\documentclass[prd,twocolumn,floatfix,superscriptaddress,showpacs,amssymb,amsmath,amsfonts,aps,letterpaper,altafillletter,nofootinbib]{revtex4}
\usepackage{color}
\usepackage{times}
\usepackage{graphicx}
\usepackage{fancyhdr}
\usepackage{float}
\usepackage{ulem}
\makeatletter
\makeatletter
\def\@fnsymbol#1{\ifcase#1\or * \or  $+$ \or  \$ \or \#  \or \dag \or \ddag \or
$\mathsection$ \or $ \mathparagraph$ \or $\|$  \or \textordfeminine \or 
\textbullet \or $\star$ \or $\clubsuit$ \or $\spadesuit$ \or \&\& \or 
\or ** \or $++$ \or  \$\$ \or \#\#  \or \dag\dag \or \ddag\ddag \or
$\mathsection\mathsection$ \or $ \mathparagraph\mathparagraph$ \or $\|\|$  \or 
\textordfeminine\textordfeminine \or \textbullet \textbullet \or *** \or $+++$ 
\or  \$\$\$ \or \#\#  \or \dag\dag \or \ddag\ddag \or
$\mathsection \mathsection\mathsection$ \or $ \mathparagraph 
\mathparagraph\mathparagraph$ \or $\|\|\|$  \or 
\textordfeminine\textordfeminine\textordfeminine \or 
\textbullet\textbullet\textbullet \or \else \@ctrerr\fi}
\makeatother

\def\thercsid{\relax}

\renewcommand{\today}{\number\day\space\ifcase\month\or
  January\or February\or March\or April\or May\or June\or
  July\or August\or September\or October\or November\or December\fi
  \space\number\year}

\begin{document}

\title{Search for gravitational waves from binary black hole inspirals in
  LIGO data}

%
%
%

\newcommand*{\AG}{Albert-Einstein-Institut, Max-Planck-Institut f\"ur Gravitationsphysik, D-14476 Golm, Germany}
\affiliation{\AG}
\newcommand*{\AH}{Albert-Einstein-Institut, Max-Planck-Institut f\"ur Gravitationsphysik, D-30167 Hannover, Germany}
\affiliation{\AH}
\newcommand*{\AN}{Australian National University, Canberra, 0200, Australia}
\affiliation{\AN}
\newcommand*{\CH}{California Institute of Technology, Pasadena, CA  91125, USA}
\affiliation{\CH}
\newcommand*{\DO}{California State University Dominguez Hills, Carson, CA  90747, USA}
\affiliation{\DO}
\newcommand*{\CA}{Caltech-CaRT, Pasadena, CA  91125, USA}
\affiliation{\CA}
\newcommand*{\CU}{Cardiff University, Cardiff, CF2 3YB, United Kingdom}
\affiliation{\CU}
\newcommand*{\CL}{Carleton College, Northfield, MN  55057, USA}
\affiliation{\CL}
\newcommand*{\CO}{Columbia University, New York, NY  10027, USA}
\affiliation{\CO}
\newcommand*{\ER}{Embry-Riddle Aeronautical University, Prescott, AZ   86301 USA}
\affiliation{\ER}
\newcommand*{\HC}{Hobart and William Smith Colleges, Geneva, NY  14456, USA}
\affiliation{\HC}
\newcommand*{\IU}{Inter-University Centre for Astronomy  and Astrophysics, Pune - 411007, India}
\affiliation{\IU}
\newcommand*{\CT}{LIGO - California Institute of Technology, Pasadena, CA  91125, USA}
\affiliation{\CT}
\newcommand*{\LM}{LIGO - Massachusetts Institute of Technology, Cambridge, MA 02139, USA}
\affiliation{\LM}
\newcommand*{\LO}{LIGO Hanford Observatory, Richland, WA  99352, USA}
\affiliation{\LO}
\newcommand*{\LV}{LIGO Livingston Observatory, Livingston, LA  70754, USA}
\affiliation{\LV}
\newcommand*{\LU}{Louisiana State University, Baton Rouge, LA  70803, USA}
\affiliation{\LU}
\newcommand*{\LE}{Louisiana Tech University, Ruston, LA  71272, USA}
\affiliation{\LE}
\newcommand*{\LL}{Loyola University, New Orleans, LA 70118, USA}
\affiliation{\LL}
\newcommand*{\MP}{Max Planck Institut f\"ur Quantenoptik, D-85748, Garching, Germany}
\affiliation{\MP}
\newcommand*{\MS}{Moscow State University, Moscow, 119992, Russia}
\affiliation{\MS}
\newcommand*{\ND}{NASA/Goddard Space Flight Center, Greenbelt, MD  20771, USA}
\affiliation{\ND}
\newcommand*{\NA}{National Astronomical Observatory of Japan, Tokyo  181-8588, Japan}
\affiliation{\NA}
\newcommand*{\NO}{Northwestern University, Evanston, IL  60208, USA}
\affiliation{\NO}
\newcommand*{\SC}{Salish Kootenai College, Pablo, MT  59855, USA}
\affiliation{\SC}
\newcommand*{\SE}{Southeastern Louisiana University, Hammond, LA  70402, USA}
\affiliation{\SE}
\newcommand*{\SA}{Stanford University, Stanford, CA  94305, USA}
\affiliation{\SA}
\newcommand*{\SR}{Syracuse University, Syracuse, NY  13244, USA}
\affiliation{\SR}
\newcommand*{\PU}{The Pennsylvania State University, University Park, PA  16802, USA}
\affiliation{\PU}
\newcommand*{\TC}{The University of Texas at Brownsville and Texas Southmost College, Brownsville, TX  78520, USA}
\affiliation{\TC}
\newcommand*{\TR}{Trinity University, San Antonio, TX  78212, USA}
\affiliation{\TR}
\newcommand*{\HU}{Universit{\"a}t Hannover, D-30167 Hannover, Germany}
\affiliation{\HU}
\newcommand*{\BB}{Universitat de les Illes Balears, E-07122 Palma de Mallorca, Spain}
\affiliation{\BB}
\newcommand*{\BR}{University of Birmingham, Birmingham, B15 2TT, United Kingdom}
\affiliation{\BR}
\newcommand*{\FA}{University of Florida, Gainesville, FL  32611, USA}
\affiliation{\FA}
\newcommand*{\GU}{University of Glasgow, Glasgow, G12 8QQ, United Kingdom}
\affiliation{\GU}
\newcommand*{\MD}{University of Maryland, College Park, MD 20742 USA}
\affiliation{\MD}
\newcommand*{\MU}{University of Michigan, Ann Arbor, MI  48109, USA}
\affiliation{\MU}
\newcommand*{\OU}{University of Oregon, Eugene, OR  97403, USA}
\affiliation{\OU}
\newcommand*{\RO}{University of Rochester, Rochester, NY  14627, USA}
\affiliation{\RO}
\newcommand*{\UW}{University of Wisconsin-Milwaukee, Milwaukee, WI  53201, USA}
\affiliation{\UW}
\newcommand*{\VC}{Vassar College, Poughkeepsie, NY 12604}
\affiliation{\VC}
\newcommand*{\WU}{Washington State University, Pullman, WA 99164, USA}
\affiliation{\WU}

\author{B.~Abbott}    \affiliation{\CT}
\author{R.~Abbott}    \affiliation{\CT}
\author{R.~Adhikari}    \affiliation{\CT}
\author{A.~Ageev}    \affiliation{\MS}  \affiliation{\SR}
\author{J.~Agresti}    \affiliation{\CT}
\author{P.~Ajith}    \affiliation{\AH}
\author{B.~Allen}    \affiliation{\UW}
\author{J.~Allen}    \affiliation{\LM}
\author{R.~Amin}    \affiliation{\LU}
\author{S.~B.~Anderson}    \affiliation{\CT}
% ...

 \collaboration{The LIGO Scientific Collaboration, http://www.ligo.org}
 \noaffiliation

\date[\relax]{ RCS \thercsid; compiled \today }
\pacs{95.85.Sz, 04.80.Nn, 07.05.Kf, 97.80.--d}

\begin{abstract}\quad
We report on a search for gravitational waves from binary black hole
inspirals in the data from the second science run of the LIGO
interferometers. The search focused on binary systems with component
masses between 3 and 20 $M_\odot$. Optimally oriented binaries with 
distances up to 1 Mpc could be detected with efficiency of at least 90\%.
We found no events that could be identified
as gravitational waves in the 385.6 hours of data that we searched.
\end{abstract}

\maketitle

\section{Introduction}
\label{intro}

The Laser Interferometric Gravitational Wave Observatory 
(LIGO) \cite{Abramovici:1992ah} consists of three 
Fabry-Perot-Michelson interferometers, which are sensitive 
to the minute changes that would be induced in the relative lengths of their 
orthogonal arms by a passing gravitational wave.
These interferometers are nearing the end of their commissioning 
phase and were close to design sensitivity as of March 2005.
During the four science runs that have been completed until now
(first (S1) during 2002, second (S2) and third (S3) during 2003 and
fourth (S4) during 2005) all three LIGO
interferometers were operated stably and in coincidence.  Although 
these science runs were performed during the commissioning phase they 
each represent the best broad-band sensitivity to gravitational 
waves that had been achieved up to that date.

In this paper we report the results of a search for 
gravitational waves from the inspiral phase of stellar mass binary 
black hole (BBH) systems, using the data from the second science run
of the LIGO interferometers.
These BBH systems are expected to emit gravitational waves at frequencies 
detectable by LIGO during the final stages of inspiral 
(decay of the orbit due to energy radiated as 
gravitational waves), the merger (rapid infall) and the subsequent 
ringdown of the quasi-normal modes of the resulting single black hole. 

The rate of BBH coalescences in the Universe is highly uncertain. 
In contrast to searches for gravitational waves from the inspiral phase
of binary neutron star (BNS) systems \cite{LIGOS2iul}, 
it is not possible to set a reliable 
upper limit on astrophysical BBH coalescences. That is because the distribution 
of the sources in space, in the component mass space and in the spin angular
momentum space is not reliably known. Additionally, the gravitational waveforms
for the inspiral phase of stellar-mass BBH systems 
which merge in the frequency band of the LIGO interferometers are not
known with precision. 
We perform a search that aims at detection of BBH inspirals. In the absence 
of a detection, we use a specific nominal model for the BBH
population in the Universe and the gravitational waveforms given in the 
literature to calculate an upper limit for the rate of BBH coalescences. 

The rest of the paper is organized as follows.
Sec.~\ref{sciruns} provides a short description of the data that 
was used for the search. In Sec.~\ref{sources} we discuss the target sources 
of the search and we explain the motivation for using
a family of phenomenological templates to search the data. 
In Sec.~\ref{filter} we give a detailed discussion
of the templates and the filtering methods. In 
Sec.~\ref{s:vetoes} we provide information on various data quality checks 
that we performed, in Sec.~\ref{pipeline} we describe in detail the
analysis method that we used and in Sec.~\ref{s:tuning} we provide details 
on the parameter tuning. In Sec.~\ref{backgr} we describe the estimation 
of the background and in Sec.~\ref{results} we present the 
results of the search. We finally show the calculation of the rate
upper limit on BBH coalescences in 
Sec.~\ref{interpretation} and we provide a brief summary of the results in
Sec.~\ref{conclusions}.


\section{Data sample}
\label{sciruns}

During the second science run, the three LIGO interferometers were
operating in science mode (see Sec.~\ref{s:vetoes}). 
The three interferometers are based at two observatories.
We refer to the observatory at Livingston, LA, as LLO and the
observatory at Hanford, WA as LHO. A total of 536 hours of data from the LLO 
4~km interferometer (hereafter L1), 1044 hours of data from the LHO 4~km 
(hereafter H1) interferometer, 
and 822 hours of data from the LHO 2~km (hereafter H2) interferometer was
obtained. The data was subjected to several quality checks.  In
this search, we used only data from times when the L1 interferometer was
running in coincidence with at least one of H1 and H2, and we only used
continuous data of duration longer than 2048 s
(see Sec.~\ref{pipeline}). After the data
quality cuts, there was a total of 101.7 hours of L1-H1 double
coincident data (when both L1 and H1 but \emph{not} H2 were operating), 
33.3 hours of L1-H2 double coincident data (when both L1 and H2 but 
\emph{not} H1 were operating) and
250.6 hours of L1-H1-H2 triple coincident data (when all three 
interferometers were operating) from the S2
data set, for a total of 385.6 hours of data.

A fraction (approximately 9\%) of this data (chosen to be representative of 
the whole run) was set aside as ``playground'' data where the various 
parameters of the analysis could be tuned and where vetoes effective 
in eliminating spurious noise events could be identified.
The fact that the tuning was performed using this subset of data does not
exclude the possibility that a detection could be made in this subset.
However, to avoid biasing the upper limit, those times were
excluded from the upper limit calculation.

As with earlier analyses of LIGO data, the output of the antisymmetric
port of the interferometer was calibrated to obtain a measure of the
relative strain $\Delta L/L$ of the interferometer arms, where
$\Delta L=L_x-L_y$ is the difference in length between the $x$ arm
and the $y$ arm and $L$ is the average arm
length.  The calibration was measured by applying known forces to the
end mirrors of the interferometers before, after 
and occasionally during the science run.  In the
frequency band between $100$~Hz and $1500$~Hz,
the calibration accuracy was within 10\% in amplitude and
$10^\circ$ of phase.


\section{Target Sources}
\label{sources}

The target sources for the search described in this paper are 
binary systems that consist of two black holes with component masses
between 3 and 20 $M_\odot$, in the last seconds before coalescence. 
Coalescences of binary systems consist of three phases: 
the inspiral, the merger and the ringdown.
We performed the search by matched filtering the data using templates
for the inspiral phase of the evolution of the binaries.
The exact duration of the inspiral signal depends on the masses of the
binary. Given the low-frequency cutoff of $100$~Hz that needed to be
imposed on the data (see Sec.~\ref{pipeline}) the expected duration 
of the inspiral signals in the S2 LIGO band as predicted
by post-Newtonian calculations varies from 0.607 s for a $3 - 3 \ M_\odot$
binary to 0.013 s for a $20 - 20 \ M_\odot$ binary.

The gravitational wave signal is dominated by the  
merger phase which potentially may be computed using numerical
solutions to Einstein's equations. Searching exclusively
for the merger using matched-filter techniques is not
appropriate until the merger waveforms are known. 
BBH mergers are usually searched for by using techniques developed for
detection of unmodeled gravitational wave bursts \cite{LIGObursts:2004}. 
However, for reasons that will be explained below, it is possible 
that the search described in this paper was also sensitive to at least
part of the merger of the BBH systems of interest.
Certain re-summation techniques have been applied to model
the late time evolution of BBH systems which makes it possible to 
evolve those systems beyond the inspiral and into the merger phase
\cite{Damour:1998zb,BuonannoDamour:1999,BuonannoDamour:2000,
DamourJaranowskiSchaefer:2000,Damour:2000zb,DamouretAl:2003,
Blanchet:2004ek} and the 
templates that we used for matched filtering incorporate the early
merger features (in addition to the inspiral phase) of those waveforms.

The frequencies of the ringdown radiation from BBH systems 
with component masses between 3 and 20 $M_\odot$ range from 295 Hz to
1966 Hz \cite{Echeverria:1989, Leaver:1985, Jolien:1999} and the
gravitational wave forms are known. Based on the frequencies of these
signals,
some of the signals are in the S2 LIGO frequency band of good sensitivity
and some are not. At the time of the search presented in this paper, the 
matched-filtering tools necessary 
to search for the ringdown phase of BBH were being developed.
In future searches we will look for ringdown signals associated with
inspiral candidates.

Finally, we have verified through simulations 
that the presence of the merger and the ringdown phases of the gravitational 
wave
signal in the data does not degrade our ability to detect the inspiral phase,
when we use matched filter techniques.

\subsection{Characteristics of BNS and BBH inspirals}

We use the standard convention $c = G = 1$ in the remainder of this paper.

The standard approach to solving the BBH evolution problem 
uses the post-Newtonian (PN) expansion \cite{Blanchet:2004ek}
of the Einstein equations to
compute the binding energy $E$ of the binary and the flux 
$F$ of the radiation at infinity, both as series expansions
in the invariant velocity $v$ (or the orbital frequency) 
of the system. This is supplemented with 
the energy balance equation ($dE/dt=-F$) which in turn 
gives the evolution of the orbital phase and hence the 
gravitational wave phase which, to the dominant order, 
is twice the orbital phase. This method works well when 
the velocities in the system are much smaller compared to
the speed of light, $v\ll 1 $. Moreover, the post-Newtonian
expansion is now complete to order $v^7$ giving us
the dynamics and orbital phasing to a high accuracy
\cite{Blanchet:2001ax,Blanchet:2002av}.
Whether the waveform predicted by the model to
such high orders in the post-Newtonian expansion
is reliable for use as a matched filter depends on how
relativistic the system is in the LIGO band. 
For the second science run of LIGO, the interferometers had
very good sensitivity between 100 and 800 Hz so we calculate
how relativistic BNS and BBH systems are at those two frequencies.

The velocity in a binary system of total mass M is
related to the frequency $f$ of the gravitational waves by
\begin{equation}
v = (\pi M f )^{1/3}.
\end{equation}
When a BNS system that
consists of two $1.4 \, M_\odot$ components enters the S2 LIGO
band, the velocity in the system is
$v \simeq 0.16 $; when it leaves the S2 LIGO band at 
800 Hz, it is $v \simeq 0.33 $ and the system is 
mildly relativistic. Thus, relativistic corrections 
are not too important for the inspiral phase of BNS.

BBH systems of high mass, however,
would be quite relativistic in the S2 LIGO band.
For instance, when a $10-10 \, M_\odot$ BBH 
enters the S2 LIGO band the velocity would be $v\simeq 0.31 $. 
At a frequency of 200 Hz (smaller than the frequency of the 
innermost stable circular orbit, explained below, which
is 220 Hz according to the test-mass approximation) the velocity
would be $v\simeq 0.40$. Such a binary is expected to merge
producing gravitational waves within the LIGO frequency band.  
Therefore LIGO would observe BBH systems in the most
non-linear regime of their evolution and thereby witness
highly relativistic phenomena for which the perturbative
expansion is unreliable. 

Numerical relativity is not yet in a position to fully
solve the late time phasing of BBH systems.
For this reason, in recent years, (non-perturbative)
analytical resummation techniques of the post-Newtonian 
series have been developed to speed up its convergence
and obtain information on the late stages of the inspiral and
the merger \cite{Blanchet:1996pi}. These resummation techniques have
been applied to the post-Newtonian expanded conservative and
non-conservative part of the dynamics and are called
effective-one-body (EOB) and P-approximants (also referred to as Pad\'e 
approximants)
\cite{Damour:1998zb,BuonannoDamour:1999,BuonannoDamour:2000,
DamourJaranowskiSchaefer:2000,Damour:2000zb,DamouretAl:2003}.
Some insights into the merger problem have been
also provided in \cite{Baker:2002qf,Baker:2001sf} 
by combining numerical and perturbative approximation schemes.

The amplitude and the phase of the standard post-Newtonian (TaylorT3,
\cite{Blanchet:1996pi}), EOB, and Pad\'e waveforms,
evaluated at different post-Newtonian orders, differ from each other
in the last stages of inspiral, close to the 
innermost stable circular orbit (ISCO, \cite{Blanchet:1996pi}).
The TaylorT3 and Pad\'e waveforms are derived assuming that the
two black holes move along a quasi-stationary sequence of circular
orbits. The EOB waveforms, extending beyond the ISCO,
contain features of the merger dynamics.
All those model-based waveforms are characterized by different ending
frequencies. For the quasi-stationary two-body models the ending frequency
is determined by the minimum of the energy. For the
models that extend beyond the ISCO,
the ending frequency is fixed by the light-ring 
\cite{BuonannoDamour:2000,BuonannoChenVallisneri:2003a} 
of the two-body dynamics.

We could construct matched filters using waveforms from each of these 
families to search for BBH inspirals but yet the true gravitational wave 
signal might be ``in between'' the models we search for. 
In order not to miss the true gravitational wave signal it is desirable to 
search a space that encompasses all the different families
and to also search the space ``in between'' them.

\subsection{Scope of the search}
Recent work by Buonanno, Chen and Vallisneri 
\cite{BuonannoChenVallisneri:2003a} 
(hereafter BCV) has unified the different 
approximation schemes into one family of phenomenological 
waveforms by introducing two new parameters, one 
of which is an amplitude correction factor and 
the other a variable frequency cutoff, in order to model
the different post-Newtonian approximations and their
variations. Additionally, in order to achieve high signal-matching 
performance, they introduced unphysical parameters
in the phase evolution of the waveform.

In this work we used a specific implementation of the phenomenological 
templates. As these phenomenological waveforms are not guaranteed to have 
a good overlap with the \emph{true} gravitational wave signal it is less
meaningful to set upper limits on either the strength of gravitational 
waves observed during our search or on the coalescence rate of BBH in the 
Universe than it was for the BNS search in the S2 data
\cite{LIGOS2iul}. However, in order to give an interpretation of the result of
our search, we did calculate an upper limit on the coalescence rate
of BBH systems, based on two assumptions: (1) that the model-based waveforms 
that exist in the literature have good overlap with a true gravitational wave 
signal and (2) that the phenomenological templates used have a good overlap 
with the majority of the model-based BBH inspiral waveforms proposed in 
the literature \cite{BuonannoChenVallisneri:2003a}.

\begin{figure}[h]
\includegraphics[width=\linewidth]{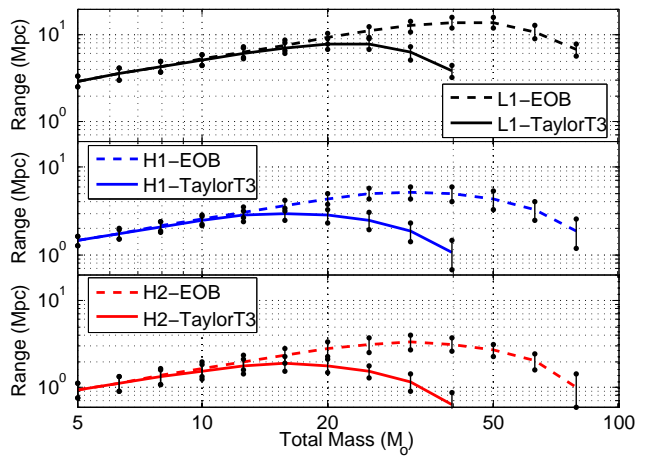}
\caption{Range (distance at which an optimally oriented
inspiraling binary of given total mass would produce a signal-to-noise
ratio of $8$) of the LIGO interferometers during S2. The error bars are
calculated from the fluctuations of the noise in the LIGO interferometers
during S2.}
\label{fig:DistanceVsMass}
\end{figure}

To set the stage for later discussion we plot in Fig.~\ref{fig:DistanceVsMass} 
the distance at which a binary of two components of equal mass
that is optimally oriented (positioned directly above the interferometer 
and with its orbital plane perpendicular to the 
line of sight from the interferometer to the binary)
would produce a signal-to-noise ratio (SNR, see Sec.~\ref{filter})
of $8$ in the LIGO interferometers during the
second science run. We refer to this distance as ``range'' of the
interferometers. 
The solid line shows the range of the LIGO interferometers 
for matched filtering performed with the standard post-Newtonian (TaylorT3)
waveforms, which predict the evolution of the system up to the ISCO
\cite{Blanchet:1996pi}, at a gravitational wave frequency of $f_{\rm GW} \sim 
110 \left (M/40 M_\odot\right )^{-1}$~Hz. 
The dashed line shows the range of the interferometers for matched filtering 
performed with the EOB
waveforms, which predict the evolution of the system up to the light ring
orbit \cite{BuonannoDamour:2000,BuonannoChenVallisneri:2003a},
at a gravitational wave frequency of $f_{\rm GW} \sim
218 \left (M / 40 M_\odot \right )^{-1}$~Hz (notice that both
these equations for $f_{\rm GW}$ are for binaries of equal component masses).
Since the EOB waveforms extend beyond the ISCO,
they have longer duration and greater energy in the LIGO band which
explains why the range for the EOB waveforms is greater than the range for
the TaylorT3 or Pad\'e waveforms (calculations performed with the Pad\'e 
waveforms result in ranges similar to those given by the TaylorT3 waveforms). 

During S2 the L1 interferometer was the most sensitive with a range of $7$ Mpc 
for a $10-10\,M_\odot$ binary (calculated using the TaylorT3 waveform). 
However, since for the search described in this paper
we demanded that our candidate events are seen in coincidence between 
the two LIGO observatories (as described in Sec.~\ref{pipeline})
the overall range of the search was determined by the less sensitive
LHO interferometers and thus was smaller than this maximum.


\section{Filtering}
\label{filter}

\subsection{Detection template family}

As was mentioned in Sec.~\ref{sources}, the gravitational wave
signal from inspiraling black hole binaries of high
masses enters the LIGO frequency sensitivity band in the
later stages, when the post-Newtonian approximation is
beginning to lose validity and different versions of the approximation are
beginning to substantially differ from each other. 
In order to detect these inspiral signals we need to use 
filters based on phenomenological waveforms (instead of model-based
waveforms) that cover the function space spanned by different versions of the
late-inspiral post-Newtonian approximation. 

It must be emphasized at this point that black hole binaries with
small component
masses (corresponding to total mass up to 10 $M_\odot$) 
enter the S2 LIGO sensitivity band at an early enough stage of the inspiral
that the signal can be adequately approximated by the stationary phase
approximation to the standard 
post-Newtonian approximation. For those binaries it is not necessary to
use phenomenological templates for the matched filtering; the standard
post-Newtonian waveforms can be used as in the search 
for BNS inspirals.
However, using the phenomenological waveforms for those
binaries does not limit the efficiency of the search 
\cite{BuonannoChenVallisneri:2003a}. In this search, in order to treat all 
black hole binaries uniformly, we chose to use the BCV templates with
parameters that span the component mass range from 3 to 20 $M_\odot$.

The phenomenological templates 
introduced in \cite{BuonannoChenVallisneri:2003a} 
match very well most physical waveform models that have been suggested
in the literature for BBH coalescences. Even though they are not derived
by calculations based on a specific physical model they are inspired by
the standard post-Newtonian inspiral waveforms. In the frequency domain, they 
are 
\begin{equation}\label{eq:template}
\tilde{h}(f) \equiv \mathcal{A}(f) e^ {i\psi(f) }, \; f > 0, 
\end{equation}
where the amplitude $\mathcal{A}(f)$ is 
\begin{equation}\label{eq:amp}
\mathcal{A}(f) \equiv f^{-7/6}\left( 1- \alpha f^{2/3}\right) 
\theta \left(f_{\textrm{cut}}-f\right)
\end{equation}
and the phase $\psi(f)$ is 
\begin{equation}\label{eq:phase}
\psi(f) \equiv 
\phi_0 + 2\pi f t_0 + f^{-5/3}\sum_{n=0}^{\infty} f^{n/3} \psi_{n}.
\end{equation}
In Eq.~(\ref{eq:amp}) $\theta$ is the Heaviside step function and
in Eq.~(\ref{eq:phase}) $t_0$ and $\phi_0$ are offsets on the time of 
arrival and on the phase of the signal respectively.  
Also, $\alpha$, $f_{\mathrm{cut}}$ and $\psi_n$ are parameters of the
phenomenological waveforms.

Two components can be identified in the amplitude part of the BCV 
templates. The $f^{-7/6}$ term comes from the restricted-Newtonian 
amplitude in the Stationary Phase Approximation (SPA) 
\cite{Droz:1999qx,thorne.k:1987,SathyaDhurandhar:1991}. 
The term $\alpha f^{2/3} \times f^{-7/6} =\alpha
f^{-1/2} $ is introduced to capture any post-Newtonian 
amplitude corrections and to give high overlaps between the BCV templates
and the various models that evolve the binary past the ISCO frequency. 
Additionally,
in order to obtain high matches with the various post-Newtonian models that
predict different terminating frequencies, a
cutoff frequency $f_{\mathrm{cut}}$ is imposed to terminate the waveform.

It has been shown \cite{BuonannoChenVallisneri:2003a} that
in order to achieve high matches with the various model-derived 
BBH inspiral waveforms
it is in fact sufficient to use only the parameters $\psi_0$ and 
$\psi_{3}$ in the phase expression in Eq.~(\ref{eq:phase}),
if those two parameters are allowed to take unphysical values. 
Thus, we set all other $\psi_n$ coefficients equal to 0
and simplify the phase to
\begin{eqnarray}
\psi(f) &=& \phi_0 + 2 \pi f t_0 + f^{-5/3} (\psi_0 + \psi_3 f)\\
& \equiv & \phi_0 +\psi_s(f)
\end{eqnarray}
where the subscript $s$ stands for ``simplified''.

For the filtering of the data, a bank of BCV templates was constructed 
over the parameters $f_{\text{cut}}$, $\psi_0$ and 
$\psi_3$ (intrinsic template parameters). For details on how the templates
in the bank were chosen see Sec.~\ref{pipeline}.
For each template, the signal-to-noise ratio (defined in Sec.~\ref{ss:filter})
is maximized over the parameters $t_0$, $\phi_0$ and $\alpha$ (extrinsic
template parameters).

\subsection{Filtering and signal-to-noise ratio maximization}
\label{ss:filter}

For a signal $s$, the signal-to-noise ratio (SNR) resulting from 
matched-filtering with a template $h$ is 
\begin{equation}
\rho(h) = \frac{\left <s,h\right >}{\sqrt{\left <h,h\right >}},
\end{equation}
with the inner product $\left <s,h\right >$ being
\begin{equation}\label{SNRDef}
\left <s,h\right > = 2 \int_{-\infty}^{\infty} 
\frac{\tilde{s}(f) \tilde{h}^{\ast}(f)}
{S_h(|f|)} df =
4 \Re \int_0^{\infty} \frac{\tilde{s}(f) \tilde{h}^{\ast}(f)}{S_h(f)} df
\end{equation}
and $S_h(f)$ being the one-sided noise power spectral density.

Various manipulations (given in detail in App.~\ref{appendixA}) give the
expression for the SNR (maximized over the 
extrinsic parameters $\phi_0$, $\alpha$ and
$t_0$) that was used in this search. That expression is
\begin{eqnarray}
\rho_{\mathrm{maximized}} &= 
 \frac{1}{2} \sqrt{|F_1|^2 +|F_2|^2+2\Im (F_1 F_2^{\ast})} \\
&+\frac{1}{2} \sqrt{|F_1|^2 +|F_2|^2 -2\Im (F_1 F_2^{\ast})}.
\nonumber
\end{eqnarray}
where
\begin{equation}
F_1 = \int_0^{f_{\mathrm{cut}}} \frac{4 \tilde{s}(f)
a_1 f^{-\frac{7}{6}}}{S_h(f)} e^{-i \psi_s(f)} df  
\end{equation}
\begin{equation}
F_2 = \int_0^{f_{\mathrm{cut}}} \frac{4 \tilde{s}(f)
(b_1 f^{-\frac{7}{6}} + b_2 f^{-\frac{1}{2}})}{S_h(f) } e^{-i \psi_s(f)} df.
\end{equation}
The quantities $a_1$, $b_1$ and $b_2$ are dependent on the noise 
and the cutoff frequency $f_{\mathrm{cut}}$ and are defined in 
App.~\ref{appendixA}.  The original suggestion of Buonanno, Chen and 
Vallisneri was that for the SNR maximization over the parameter $\alpha$ 
the values of $(\alpha \times f_{\mathrm{cut}}^{2/3})$
should be restricted within the range $[ 0, \ 1]$, for reasons
that will be explained in Sec.~\ref{sss:singleifotuning}. However, in order to 
be able to perform various investigations on the values of $\alpha$ we
leave its value unconstrained in this maximization procedure. More
details on this can be found in Sec.~\ref{sss:singleifotuning}.



\section{Search for events}

\subsection{Data quality and veto study}
\label{s:vetoes}

The matched filtering algorithm is optimal for data with a known
calibrated noise spectrum that is Gaussian and stationary over the time scale 
of the data blocks analyzed ($2048$~s, described in
Section~\ref{pipeline}), which requires stable,
well-characterized interferometer performance.  In practice, the
performance is influenced by non-stationary optical alignment, servo
control settings, and environmental conditions.  We used two
strategies to avoid problematic data. The first strategy was to evaluate data
quality over relatively long time intervals using several different
tests. As in the BNS search, 
time intervals identified as being unsuitable for analysis
were skipped when filtering the data.  The second strategy was to look
for signatures in environmental monitoring channels and auxiliary
interferometer channels that would indicate an environmental disturbance
or instrumental transient,  allowing us to
veto any candidate events recorded at that time. 

The most promising candidate for a veto channel was L1:LSC-POB\_I 
(hereafter referred to as ``POBI''), an auxiliary channel measuring signals 
proportional to the length fluctuations of the power recycling cavity. 
This channel was found to have highly variable noise at $70$~Hz which 
coupled into the gravitational wave channel. Transients found in this channel 
were used as vetoes for the BNS search in the S2 data 
\cite{LIGOS2iul}. Hardware injections of simulated inspiral signals
\cite{S2Hardware:2004} were used to 
prove that signals in POBI would not veto true inspiral gravitational waves
present in the data. 



Investigations showed that using the correlations between POBI and the
gravitational wave channel to veto candidate events would be less
efficacious than it was in the BNS search.
Therefore POBI was not used a an a-priori veto.
However, the fact that correlations were proven to 
exist between the POBI signals and the BBH inspiral signals made 
it worthwhile to follow-up the BBH inspiral events 
that resulted at the end of our analysis
and check if they were correlated with POBI signals (see 
Sec.~\ref{pobicorrelations}).

As in the BNS search in the S2 data, no instrumental vetoes were found
for H1 and H2.
A more extensive discussion of the LIGO S2 binary inspiral veto studies can be
found in \cite{vetoGWDAW03}.

\subsection{Analysis Pipeline}
\label{pipeline}

In order to increase the confidence that a candidate
event coming out of our analysis is a true gravitational wave and not due to
environmental or instrumental noise we demanded the candidate event to be
present in the L1 interferometer and at least one of the LHO interferometers.
Such an event would then be characterized as a potential 
inspiral event and be subject to thorough examination.

The analysis pipeline that was used to perform the BNS
search (and was described in detail in \cite{LIGOS2iul}) was the starting
structure for constructing the 
pipeline used in the BBH inspiral search described in this paper.
However, due to the different nature of the search, the details of
some components of the pipeline needed to be modified.
In order to highlight the differences of the two pipelines and to explain 
the reasons for those, we describe our pipeline below.  

First, various
data quality cuts were applied on the data and the segments of good data for
each interferometer were indentified.
The times when each interferometer was in stable operation (called science
segments) were used to construct three data sets corresponding to: 
(1) times when all three interferometers were operating (L1-H1-H2
triple coincident data),
(2) times when \emph{only} the L1 and H1 (and \emph{not} the H2)
interferometers were operating (L1-H1 double coincident data) and
(3) times when \emph{only} the L1 and H2 (and \emph{not} the H1)
interferometers were operating (L1-H2 double coincident data).
The analysis pipeline produced a list of coincident triggers (times 
and template parameters for which
the SNR threshold was exceeded and all cuts mentioned below
were passed) for each of the three data sets.

The science segments were analyzed in blocks of 2048 s
using the \textsc{findchirp} implementation \cite{findchirppaper}
of matched filtering for
inspiral signals in the LIGO Algorithm library~\cite{LALS2BBH}. The
original version of \textsc{findchirp} had been coded for the BNS
search and thus had to be modified to allow filtering of the
data with the BCV templates described in Sec.~\ref{filter}. 

The data for each 2048 s block was first down-sampled from 16384~Hz to
4096~Hz.  It was subsequently high-pass filtered at 90~Hz in the
time domain and a low frequency cutoff of 100~Hz was imposed in the 
frequency domain.  The instrumental response for
the block was calculated using the average value of the calibration
(measured every minute) over the duration of the block.

The breaking up of each segment for power spectrum estimation and for
matched-filtering was identical to the BNS search \cite{LIGOS2iul} and is 
briefly mentioned here so that the terminology is established for the 
pipeline description that follows.

\begin{figure}
\begin{center}
\hspace*{-0.2in}\includegraphics[width=\linewidth]{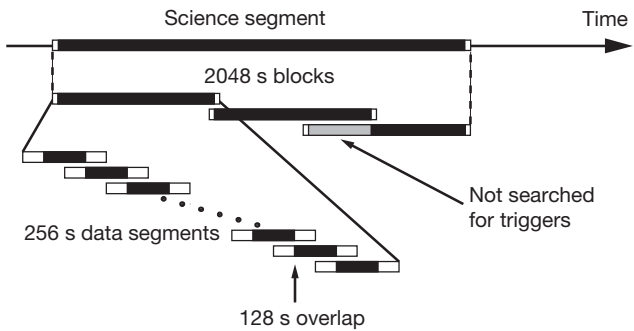}
\end{center}
\caption{\label{f:chunks}%
The algorithm used to divide science segments into data analysis
segments.  Science segments are divided into $2048$~s blocks
overlapped by $128$~s. (Science segments shorter than $2048$~s are
ignored.) An additional block with a larger overlap is added to cover
any remaining data at the end of a science segment.  Each block is
divided into $15$ analysis sesgments of length $256$~s for
filtering. The first and last $64$~s of each analysis segment are
ignored, so the segments overlap by $128$~s.  Areas shaded black are
searched for triggers by the search pipeline. The gray area in the
last block of the science segment is not searched for triggers as this
time is covered by the preceding block, although these data points are
used in estimating the noise power spectral density for the
final block.}
\end{figure}

Triggers were not searched for
within the first and last $64$~s of a given block, so subsequent
blocks were overlapped by $128$~s to ensure that all of the data in a
continuous science segment (except for the first and last 64
s) was searched.  Any science segments shorter than
$2048$~s were ignored. If a science segment could not be exactly divided
into overlapping blocks (as was usually the case) the remainder of the
segment was covered by a special $2048$ s block which overlapped with the
previous block as much as necessary to allow it to reach the end of
the segment.  For this final block, a parameter was set to restrict the
inspiral search to the time interval not covered by any previous block,
as shown in Fig.~\ref{f:chunks}.

Each block was further split into 15 analysis segments of length 256
s overlapped at the beginning and at the end by 64 s.  The average power
spectrum $S_h(f)$ for the 2048 s of data was estimated by taking
the median of the power spectra of the 15 segments.  We used the
median instead of the mean to avoid biased estimates due to large
outliers, produced by non-stationary data. The 
calibration was applied to the data in each analysis segment. 

In order to avoid end-effects due to wraparound of the
discrete Fourier transform when performing the matched filter, the
frequency-weighting factor $1/S_h(f)$ was truncated in the time domain so that 
its inverse Fourier transform had a maximum duration of $\pm 16$ s.
The output of the matched filter near the beginning and end of each
segment was corrupted by end-effects due to the finite duration of the
power spectrum weighting and the template.  By ignoring
the filter output within 64~s of the beginning and end of each segment,
we ensured that only uncorrupted filter output is searched for inspiral
triggers.  This necessitated the overlapping of segments and blocks 
described above.

The single-sided power spectral density (PSD) of the noise 
$S_h(f)$ in the L1 interferometer
was estimated independently for each L1 block that was
coincident with operation of at least one LHO interferometer. 
The PSD was used to construct a
template bank for filtering that block. The bank was constructed 
over the parameters $\psi_0$ and $\psi_3$ so that
there was no less than $95\%$ overlap (defined in the sense of 
Eq.~(\ref{overlap2}) between two neighboring
(in $\{ \psi_0,\ \psi_3,\ f_{\mathrm{cut}} \} $ parameter space) templates, 
if the value of $\alpha$ of those templates was equal to 0.
The $\psi_0-\psi_3$ space was tiled using a square grid based on the metric
in Eq.~(117) of \cite{BuonannoChenVallisneri:2003a}.
For each pair of $\psi_0$ and $\psi_3$, three values of the cutoff
frequency $f_{\mathrm{cut}}$ were generated. This process gave the three
intrinsic parameters for each template. Details on the
exact values of the parameters used are given in Sec.~\ref{sss:singleifotuning}.
The number of templates in the bank varied with
the PSD. For this search the number of templates ranged between
741 and 1296 templates per 2048 s L1 analysis block, with the average number
being 958.

The data from the L1 interferometer for the block was then matched-filtered, 
as described previously, against 
the bank of templates, with a SNR threshold 
$\rho_{\mathrm{L}}^{\mathrm{thresh}}$ to produce a list of triggers. As
will be explained below, the $\chi^2$-veto \cite{Allen:2004}
that was used for the BNS search was not used in this search.
For each block in the LHO interferometers, a {\em triggered bank} was
created consisting of every template that produced at least one
trigger in the L1 data during the time of the LHO block.  This 
triggered bank was used to matched-filter the data from the LHO 
interferometers. For times when only the H2
interferometer was operating in coincidence with L1, the triggered bank was
used to matched-filter the H2 blocks that overlapped with L1 data. 
For all other times, all H1 data that overlapped with L1 data was 
matched-filtered
using the triggered bank for that block. For H1 triggers produced
during times when all three interferometers were operating, a second
triggered bank was produced for each H2 block. This
triggered bank consisted of every
template which produced at least one trigger found in coincidence in
L1 and H1 during the time of the H2 block. The H2 block was
matched-filtered with this bank.  

Before any triggers were tested for coincidence, all triggers with 
$(\alpha \times f_{\text{cut}}^{2/3})$  greater than a threshold
$\alpha_F^{\text{thresh}}$ were rejected. The reason for this veto
will be explained in Sec. \ref{s:tuning}.

For triggers to be considered coincident between two interferometers they
had to be observed in both interferometers within a time window
that allowed for the error in measurement of the time of the trigger.
If the interferometers 
were not co-located, this parameter was increased by the
light travel time between the two LIGO observatories ($10$~ms). 
We then ensured that
the triggers had consistent waveform parameters by demanding that the two 
parameters $\psi_0$ and $\psi_3$ for the template were exactly equal to each
other.

Triggers that were generated from the triple coincident data
were required to be found in coincidence in the L1 and H1 
interferometers. We searched the H2 data from these triple coincident times
but did not reject L1-H1 coincident triggers that were not found in the H2 
data, even if, based on
the SNR observed in H1 they would be expected to be
found in H2. This was a looser rejection algorithm than the one used for
the BNS search \cite{LIGOS2iul}
and could potentially increase the number of 
false alarms in the triple coincident data.
The reason for using this algorithm in the BBH search is explained in
Sec.~\ref{coinctune}, where the coincidence parameter tuning is
discussed.

The last step of the pipeline was the clustering of the triggers. The 
clustering is necessary because both large astrophysical signals and
instrumental noise bursts can produce many triggers with coalescence
times within a few ms of each other and with different template 
parameters. Triggers separated by more than 0.25 s were considered distinct. 
This time was approximately half of the duration of the longest
signal that we could detect in this search. We chose the trigger with the 
largest combined SNR from each cluster, where the combined SNR is defined 
in Sec.~\ref{backgr}.

To perform the search on the full data set, a Directed Acyclic Graph (DAG) was
constructed to describe the work flow, and execution of the pipeline tasks
was managed by Condor~\cite{beowulfbook-condor} on the various Beowulf
clusters of the LIGO Scientific Collaboration. 
The software to perform all steps of the analysis pipeline and construct the
DAG is available in the package \textsc{lalapps} \cite{LALS2BBH}.

\subsection{Parameter Tuning}
\label{s:tuning}

An important part of the analysis was to decide on the values of 
the various parameters of the search, such as the SNR
thresholds and the coincidence parameters. 
The parameters were chosen so as to compromise between increasing the 
detection efficiency and lowering
the number of false alarms. 

The tuning of all the parameters was done by studying the playground data only.
In order to tune the
parameters we performed a number of Monte-Carlo simulations, in which we 
added simulated BBH inspiral signals in the data and searched for them with our
pipeline. While we used the phenomenological 
detection templates to perform the matched filtering, 
we used various model-based waveforms for the simulated signals
that we added in the data. Specifically, we chose to inject
effective-one-body (EOB, 
\cite{Damour:1998zb,BuonannoDamour:1999,BuonannoDamour:2000,Damour:2000zb}), 
Pad\'e (Pad\'eT1, \cite{Damour:1998zb})
and standard post-Newtonian waveforms (TaylorT3, \cite{Blanchet:1996pi}),
all of second post-Newtonian order. Injecting waveforms from different families
allowed us to 
additionally test the efficiency of the BCV templates for recovering signals
predicted by different models.

In contrast to neutron stars, there are no observation-based
predictions about the population of BBH systems in the Universe.
For the purpose of tuning the parameters of our pipeline 
we decided to draw the signals to be added in the data from a 
population with distances between 10 kpc and 20 Mpc from the Earth. 
The random sky positions and orientations of the binaries
resulted in some signals having much larger effective distances
(distance from which the binary would give the same signal in the
data if it were optimally oriented).
It was determined that using a uniform-distance or uniform-volume
distribution for the binaries would overpopulate the larger distances
(for which the LIGO interferometers were not very sensitive during S2)
and only give a small number of signals in the small-distance region,
which would be insufficient for the parameter tuning. For that reason
we decided to draw the signals from a population that was uniform in
$\log$(distance). For the mass distribution, we limited each 
component mass between 3 and 20 $M_\odot$. Populations 
with uniform distribution of total mass were 
injected for the tuning part of the analysis.

There were two sets of parameters that we could tune in the pipeline: 
the single interferometer parameters, which were used in the matched filtering 
to generate inspiral triggers in each interferometer, and the
parameters which were used to determine if triggers from 
different interferometers
were coincident. The single interferometer parameters that needed to be
tuned were the ranges of values for $\psi_0$ and $\psi_3$ in the template
bank, the number of $f_\mathrm{cut}$ frequencies for each pair of
$\{ \psi_0 , \psi_3 \}$ in the template bank and 
the SNR threshold $\rho^\mathrm{thresh}$. 
The coincidence parameters were the time
coincidence window for triggers from different interferometers, 
$\delta t$, and the coincidence window for
the template parameters $\psi_0$ and $\psi_3$. Due to the nature of the 
triggered search pipeline, parameter tuning was carried out in two stages. We
first tuned the single interferometer parameters for the primary interferometer
 (L1).
We then used the triggered template banks (generated from the L1 triggers)
to explore the single
interferometer parameters for the less sensitive LHO interferometers. Finally
the parameters of the coincidence test were tuned.

\subsubsection{Single interferometer tuning}
\label{sss:singleifotuning}

Based on the playground injection analysis it was determined that
the range of values for $\psi_0$ had to be 
$[ 10 ,\ 550000 ] \ \mathrm{Hz}^{5/3}$ 
and the range of values for $\psi_3$ had to be
$[ -4000,\ -10 ] \ \mathrm{Hz}^{2/3}$
in order to have high detection efficiency for  
binaries of total mass between 6 and 40 $M_\odot$. 

Our numerical studies showed that using between 3 and 5 cutoff frequencies per 
$\{\psi_0,\ \psi_3\}$ pair would yield
very high detection efficiency. Consideration of the computational
cost of the search led us to use 3 cutoff frequencies per pair, thus
reducing the number of templates in each bank by 40\% compared to
a template bank with 5 cutoff frequencies per $\{\psi_0,\ \psi_3\}$ pair. 

Our Monte-Carlo simulations showed that, in order to be able to distinguish
an inspiral signal from an instrumental or environmental
noise event in the data, the minimum
requirement should be that the trigger has SNR of at least
$7$ in each interferometer. 
A threshold of 6 (that was used in the BNS search) resulted in a
very large number of noise triggers that needlessly complicated
the data handling and post-pipeline processing.

A standard part of the matched-filtering process is the $\chi^2$-veto 
\cite{Allen:2004}. 
The $\chi^2$-veto compares the SNR accumulated in each of a number of
frequency bands of equal inspiral template power to the expected
amount in each band. Gravitational waves from inspiraling binaries give
small $\chi^2$ values while instrumental artifacts give high $\chi^2$ values.
Thus, the triggers resulting from instrumental artifacts can be vetoed
by requiring the value of $\chi^2$ for a trigger to be below a threshold.
The test is very efficient at distinguishing 
BNS inspiral signals from loud non-Gaussian noise events in the
data and was used in the BNS inspiral search \cite{LIGOS2iul} in the 
S2 data. However, we found that the $\chi^2$-veto  was 
not suitable for the search for gravitational
waves from BBH inspirals in the S2 data.
The expected short duration, low bandwidth and
small number of cycles in the S2 LIGO frequency band for many of the
possible BBH inspiral signals made such a test unreliable unless
a very high threshold on the values of $\chi^2$ were to be set. 
A high threshold, on the other hand,
resulted in only a minimal reduction in the number of noise
events picked up. Additionally the
$\chi^2$-veto is computationally very costly. 
We thus decided to not use it in this search.

\begin{figure}[ht]
\begin{center}
\includegraphics[width=\linewidth]{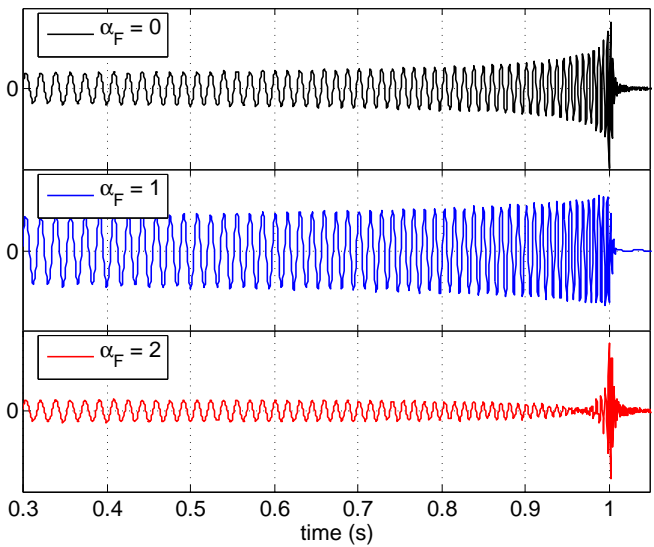}
\end{center}
\caption{\label{f:alphaF}
Time domain plots of BCV waveforms for different values of $\alpha_{F}$. The top
plot is for $\alpha_F = 0$, the middle is for $\alpha_F = 1$ and
the bottom is for $\alpha_F = 2$. For all three waveforms 
$\psi_0 = 150000 \ \mathrm{Hz}^{5/3}$, $\psi_3 = -1500 \ \mathrm{Hz}^{2/3}$ 
and $f_{\mathrm{cut}} = 500$~Hz. It can be seen that the 
behavior is not that of a typical inspiral waveform for $\alpha_F = 2$.}
\end{figure}

As mentioned earlier, the SNR calculated using the BCV templates
was maximized over the template parameter $\alpha$.
For every value of $\alpha$, there is a frequency $f_0$ for which the 
amplitude factor $(1 - \alpha f^{2/3})$ becomes zero:
\begin{equation}
f_0 = \alpha^{-3/2}.
\end{equation}
If the value of $\alpha$ associated with a trigger is such that the frequency 
$f_0$ is greater than the cutoff frequency $f_{\mathrm{cut}}$ of the template
(and consequently $\alpha \ f_{\mathrm{cut}}^{2/3} \leq 1$),
then the high-frequency behavior of the phenomenological template 
is as expected for an inspiral gravitational waveform.
If the value of $\alpha$ is such that $f_0$ is smaller than the cutoff
frequency $f_{\mathrm{cut}}$ (and consequently $\alpha \ 
f_{\mathrm{cut}}^{2/3} > 1$), the
amplitude of the phenomenological waveform becomes zero before the cutoff 
frequency is reached. For such a waveform, the high-frequency behavior does not
resemble that of a typical inspiral gravitational waveform. 
For simplicity we define
\begin{equation}
\alpha_F \equiv \alpha f_{\mathrm{cut}}^{2/3} .
\end{equation}
The behavior of the BCV waveforms for three different values of $\alpha_F$ is
shown in Fig.~\ref{f:alphaF}, where it can be seen that for the case
of $\alpha_F = 2$ the amplitude becomes zero and then increases again.

Despite that fact, many of the simulated signals that we added in the data 
\emph{were} in fact recovered
with values of $\alpha_F > 1$, with a higher SNR than the SNR
they would have been recovered with, had we imposed a restriction on $\alpha$.
Additionally some signals gave SNR smaller than
the threshold for all values of $\alpha$ that gave $\alpha_F \leq 1$.
Multiple studies showed that this was
due to the fact that we only had a limited number of cutoff frequencies
in our template bank and in many cases the lack of the appropriate 
ending frequency was compensated for by a value of $\alpha$ that corresponded
to an untypical inspiral gravitational waveform. 

We performed various investigations which showed that rejecting triggers 
with $\alpha_F > 1$ 
allowed us to still have a very high efficiency in detecting BBH inspiral
signals (although not as high as if we did not impose that cut)
and the cut primarily affected signals that were recovered with SNR close to
the threshold. It was also proven that such a cut 
reduced the number of noise triggers significantly, 
so that the false-alarm probability was significantly
reduced as well. The result was that 
such a cut provided a clearer distinction between
the noise triggers that resulted from our pipeline and the triggers
that came from simulated signals injected in the data.
In order to increase our confidence in the triggers that came out of the 
pipeline being BBH inspiral signals, 
we rejected all triggers with $\alpha_F > 1$ in this search.

As mentioned in Sec.~\ref{filter}, the initial suggestion of Buonanno,
Chen and Vallisneri was that the parameter $\alpha_F$ be constrained from
below to not take values less than zero. This suggestion was based on the
fact that for values of $\alpha < 0$, the amplitude factor 
$( 1 - \alpha f_{\mathrm{cut}}^{2/3} )$ can substantially deviate from
the predictions of the post-Newtonian theory at high frequencies.
Investigations similar to those described for the cut $\alpha_F \le 1$
did not justify rejecting the triggers with $\alpha_F < 0$, so we set
no low threshold for $\alpha_F$.

\subsubsection{Coincidence parameter tuning}
\label{coinctune}

After the single interferometer parameters had been selected, the coincidence
parameters were tuned using the triggers from the single interferometers.

The time of arrival of a simulated signal at an interferometer 
could be measured within $\pm 10$~ms. Since the H1 and H2
interferometers are co-located, $\delta t$ was chosen to be $10$~ms for
H1-H2 coincidence. Since the light-travel time
between the two LIGO observatories is $10$~ms, $\delta t$ was chosen to be 
$20$~ms for LHO-LLO coincidence. 

Because we performed a triggered search, the data from all three LIGO
interferometers 
was filtered with the same templates for each 2048 s segment. 
That led us to set the values
for the template coincidence parameters $\Delta \psi_0$ and $\Delta \psi_3$ 
equal to $0$. We found that that was sufficient for the simulated BBH
inspiral signals to be recovered in coincidence.

The slight misalignment of the L1 interferometer with respect to the 
LHO interferometers led us to choose to not impose an amplitude cut in 
triggers that came from the two different observatories. 
This choice was identical to the
choice made for the BNS search \cite{LIGOS2iul}.

We considered imposing an amplitude
cut on the triggers that came from triple coincident data and were
otherwise coincident between H1 and H2. A similar cut was imposed on
the equivalent triggers in the BNS search. The cut relied on the calculation of
the ``BNS range'' for H1 and H2. The BNS range is
defined as the distance at which an 
optimally oriented neutron star binary, consisting of two components
each of 1.4 $M_\odot$, would be detected with a SNR of 8 
in the data. The value of the range depends on the
PSD. 
For binary neutron stars 
the value of the range can be calculated for the 1.4-1.4 $M_\odot$ binary
and then be rescaled for all masses.
That is because the BNS inspiral signals always terminate
at frequencies above 733 Hz (ISCO frequency for a 3-3 $M_\odot$ 
binary, according to the test-mass approximation) 
and thus for BNS the larger part of the SNR comes from the 
high-sensitivity band of LIGO.
For black hole binaries, on the other hand, the ending frequency of
the inspiral varies from 110 Hz for a 20-20 $M_\odot$ binary up to 733 Hz
for a 3-3 $M_\odot$ binary (according to the test-mass approximation).
That means that the range depends not only on the PSD
but also on the binary that is used to calculate it. That can make 
a cut based on the range very unreliable and force rejection
of triggers that should not be rejected. In order to be sure that we
would not miss any BBH inspiral signals, we decided to not impose the
amplitude consistency cut between H1 and H2.



\section{Background Estimation}
\label{backgr}

We estimated the rate of accidental coincidences
(also known as background rate) for this search by introducing an artificial
time shift $\Delta t$ to the triggers coming from the
L1 interferometer relative to the LHO interferometers.  The time-shift
triggers were fed into the coincidence steps of the pipeline and,
for the triple coincident data, to the step of the filtering of the H2 data
and the H1-H2 coincidence. By choosing a shift larger than
20~ms (the time coincidence window between the two observatories),
we ensured that a true gravitational wave could never produce coincident
triggers in the time-shifted data streams. To avoid correlations, we used
shifts longer than the duration of the longest waveform that we
could detect (0.607 s given the low-frequency cutoff imposed, as explained
in Sec.~\ref{sources}).
We chose to not time-shift the data from the two LHO interferometers
relative to one
another since there could be true correlations producing accidental
coincidence triggers due to environmental disturbances affecting both of them.
The resulting time-shift triggers 
corresponded only to accidental coincidences of noise triggers.
For a given time shift, the triggers that emerged from the pipeline were
considered as one single trial representation of an output from a search if
no signals were present in the data.

A total of 80 time-shifts were performed and analyzed 
in order to estimate the background. The time shifts ranged from
$\Delta t = - 407\; s$ up to $\Delta t = + 407 \;s$ in increments of
$10 \; s$. The time shifts of $\pm 7 s$ were not performed.

\begin{figure}[ht]
\begin{center}
\includegraphics[width=\linewidth]{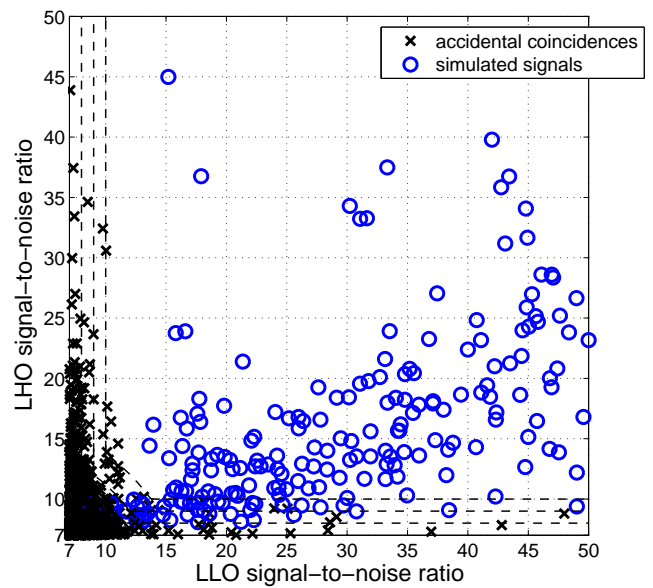}
\end{center}
\caption{\label{f:background}
The LLO and LHO
SNRs of the accidental coincidences from the time-shifted triggers (crosses)
and the triggers from the simulated signal injections (circles)
are shown. The dashed lines show the equal false-alarm contours.}
\end{figure}

\subsection{Distribution of background events}

Fig.~\ref{f:background} shows the time shift triggers that resulted from 
our pipeline (crosses) in the LHO SNR $(\rho_{\mathrm{H}})$ versus 
the LLO SNR $(\rho_{\mathrm{L}})$ plane. 
There were only double coincident triggers in both the double and triple 
coincident data. Specifically, all the triggers present in the triple
coincident data were L1-H1 coincident and were not seen in H2. Thus,
$\rho_{\mathrm{H}}$ is defined as the SNR of either H1 or H2,
depending on which of the three S2 data sets described in Sec.~\ref{pipeline}
the trigger came from.

There is a 
group of triggers at the lower left corner of the plot, which correspond
to coincidences with SNR of no more than $15$ in each observatory. 
There are also long ``tails'' of triggers which have high SNR
(above 15) in one of the interferometers and low SNR
(below 10) in the other. The distribution is quite different from
the equivalent distribution that was observed in the BNS search
\cite{LIGOS2iul}. The presence of the tails
(and their absence from the corresponding distribution in the BNS
search) can be attributed
to the fact that the $\chi^2$-veto was not applied in this search
and thus some of the loud noise events that would have been
eliminated by that test have instead survived.

For comparison, the triggers from some recovered injected signals are
also plotted in Fig.~\ref{f:background} (circles). The 
distribution of those triggers is quite different from the distribution
of the triggers resulting from accidental coincidences of noise events.
The noise triggers are concentrated along the two axes
of the ($\rho_L, \ \rho_H$) plane and the injection triggers are
spread in the region below the equal-SNR diagonal line of the same plane. 
This distribution of the injection triggers is due to the fact that 
during the second science run the L1 interferometer was more 
sensitive than the LHO interferometers. Consequently, a 
true gravitational wave signal that
had comparable LLO and LHO effective distances would be observed with
higher SNR in LLO than in LHO. The few injections that produced
triggers above the diagonal of the graph correspond to BBH 
systems that are better oriented for the LHO interferometers 
than for the L1 interferometer, and thus have higher SNR
at LHO than at LLO.

\subsection{Combined SNR}

We define a ``combined SNR'' for the coincident triggers that come out of the 
time-shift analysis. The combined SNR is a statistic
based on the accidental coincidences
and is defined so that the higher it is for a trigger, the less likely 
it is that the trigger is due to an accidental coincidence of 
single-interferometer uncorrelated noise triggers.
Looking at the plot of the $\rho_{\mathrm{H}}$ versus the $\rho_{\mathrm{L}}$
of the background
triggers we notice that the appropriate contours for the triggers at the
lower left corner of the plot are concentric circles with the
center at the origin. However, for the tails along the axes the appropriate
contours are ``L'' shaped. The combination of those two kinds of contours
gives the contours plotted with dashed lines in Fig.~\ref{f:background}. 
Based on these contours, we define the combined SNR of a trigger to be
\begin{equation}
\rho_{\mathrm{C}} = \min\{ \sqrt{\rho_{\mathrm{L}}^2 + \rho_{\mathrm{H}}^2},\;
 2 \rho_{\mathrm{H}} - 3,\; 2 \rho_{\mathrm{L}} - 3\}.
\label{e:combsnr}
\end{equation}
After the combined SNR is assigned to each pair of triggers, 
the triggers are clustered
by keeping the one with the
highest combined SNR within 0.25 s, thus keeping the ``highest
confidence'' trigger for each event.

\section{Results}
\label{results}

In this section we present the results of the search in the S2 data with the
pipeline described
in Sec.~\ref{pipeline}. The combined SNR was assigned to the
candidate events according to Eq.~(\ref{e:combsnr}). 

\subsection{Comparison of the unshifted triggers to the background}
There were 25 distinct candidate
events that survived all the analysis cuts. Of those, 7 
were in the L1-H1 double coincident data, 10 were in the L1-H2 double 
coincident data and 8 were in the L1-H1-H2 triple coincident data. Those
8 events appeared only in the L1 and H1 data streams and even though
they were not seen in H2 they were still kept, according to the procedure
described in Sec.~\ref{pipeline}.

\begin{figure}[ht]
\begin{center}
\includegraphics[width=\linewidth]{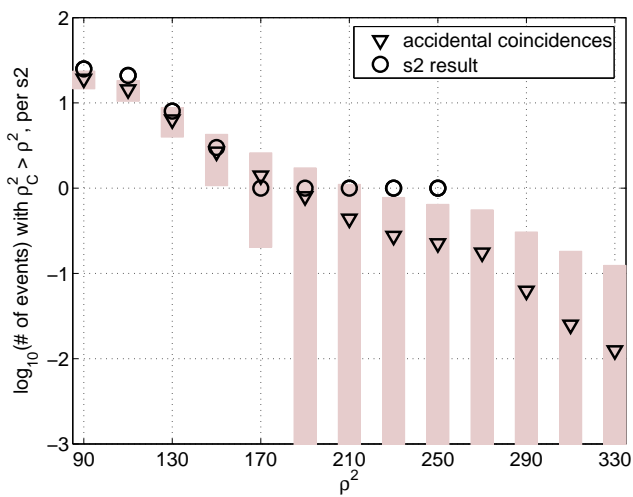}
\end{center}
\caption{\label{f:eventsPerS2}
Expected accidental coincidences per S2 (triangles) with one 
standard deviation bars. 
The number of events in the S2 (circles) is overlayed.}
\end{figure}

In order to determine if there was an excess of candidate events above the 
background in the S2 data, we compared the number of zero-shift events 
to the expected number of
accidental coincidences in S2, as predicted by the time-shift 
analysis described 
in Sec.~\ref{backgr}. Fig.~\ref{f:eventsPerS2} shows the mean
cumulative number of accidental coincidences (triangles) versus the combined
SNR squared of those accidental coincidences. 
The bars indicate one standard deviation.
The cumulative number of candidate events in the zero-shift
S2 data is overlayed (circles).
It is clear that the candidate events are consistent with the background.

\subsection{Investigations of the zero-shift candidate events}
\label{pobicorrelations}

Even though the zero-shift candidate events are consistent with the background, 
we investigated them carefully. We first looked at the possibility of those
candidate events being correlated with events in the POBI channel, 
for the reasons described in Sec.~\ref{s:vetoes}.

It was determined that the loudest candidate event and three of the remaining 
candidate events that resulted from our analysis were  
coincident with noise transients in POBI.
That led us to believe that the source of these 
candidate events was instrumental
and that they were not due to gravitational waves.
The rest of the candidate events were indistinguishable from the background
events.

\subsection{Results of the Monte-Carlo simulations}

As mentioned in Sec.~\ref{pipeline}, the Monte-Carlo simulations allow us
not only to tune the parameters of the pipeline, but also to measure the
efficiency of our search method. In this section we look in detail at the
results of Monte-Carlo simulations in the full data set of the second science
run.

Due to the lack of observation-based predictions for the population of
BBH systems in the Universe,
the inspiral signals that we injected were uniformly distributed in 
$\log$(distance), with distance varying from 10 kpc to 20 Mpc and
uniformly distributed in component mass 
(this mass distribution was proposed by \cite{Belczynski:2002}),
with each component mass varying between 3 and 20 $M_\odot$.

\begin{figure}[ht]
\begin{center}
\includegraphics[width=\linewidth]{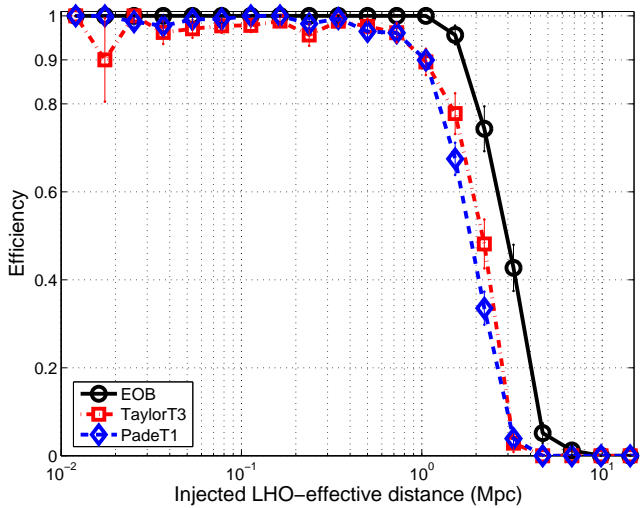}
\end{center}
\caption{\label{f:effic}
The efficiency versus the LHO-effective distance for the different
families of injected waveforms is shown. The dashed line represents the
efficiency for Pad\'eT1 injections, the dotted line represents the
efficiency for standard post-Newtonian time-domain waveforms and the
solid line represents the efficiency for effective-one-body waveforms.
All injected waveforms were of second post-Newtonian order.
Binomial error bars are shown.}
\end{figure}

Fig.~\ref{f:effic} shows the efficiency of recovering the injected signals 
(number of found injections of a given effective distance divided by the total
number of injections of that effective distance) 
versus the injected LHO-effective distance.
We chose to plot the efficiency versus the LHO-effective distance rather
than versus the LLO-effective distance since H1 and H2 were less sensitive
than L1 during S2.

Our analysis method had efficiency of at least $90 \%$ for
recovering BBH inspiral signals with LHO-effective distance less than 1 Mpc
for the mass range we were exploring.
It should be noted how the efficiency of our pipeline varied 
for different injected waveforms.
It is clear from Fig.~\ref{f:effic} that the efficiency for
recovering EOB waveforms was higher than that for
TaylorT3 or Pad\'eT1 waveforms for all distances. 
This is expected because the EOB waveforms have more power
(longer duration and larger number of cycles) in the LIGO frequency band
compared to the Pad\'eT1 and TaylorT3 waveforms.

\begin{figure}[ht]
\begin{center}
\includegraphics[width=\linewidth,clip]{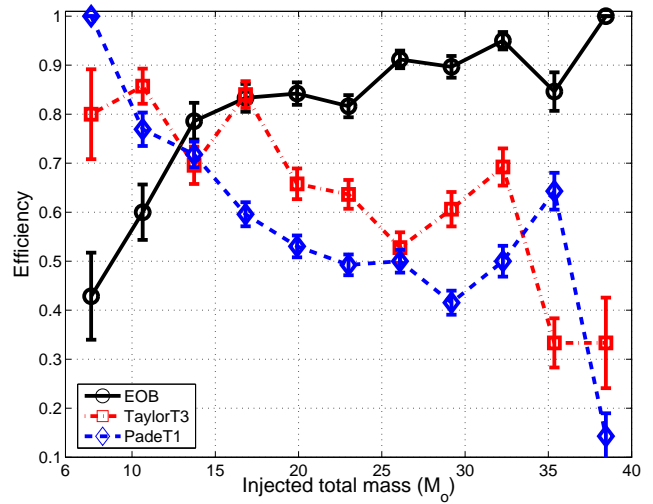}
\end{center}
\caption{\label{f:efficvsM}
The efficiency versus the injected total mass of all injected
signals with LHO-effective distance between 1 and 3 Mpc for the different
families of injected waveforms is shown. The dashed line represents the
efficiency for Pad\'eT1 injections, the dotted line represents the
efficiency for standard post-Newtonian time-domain waveforms and the
solid line represents the efficiency for effective-one-body waveforms.
All injected waveforms were of second post-Newtonian order.
Binomial error bars are shown.}
\end{figure}

Even though the main determining factor of whether a simulated signal is 
recovered or not is its effective distance in the least sensitive
interferometer, it is worth investigating the 
efficiency of recovering injections as a function of the injected total
mass. In order to limit the effect of the distance in the efficiency, we
chose all the injections with LHO-effective distance between 1 and 3 
Mpc and plotted the efficiency versus total mass in Fig.~\ref{f:efficvsM}.
The plot shows a decrease in the detection efficiency as the
total mass increases, for the TaylorT3 and the 
Pad\'eT1 waveforms, but not for the EOB waveforms.
The reason for this is the same as the one mentioned previously. 
For the TaylorT3 and the Pad\'eT1 waveforms, the
higher the total mass of the binary system, the fewer the cycles in the LIGO
band. That leads to the reduction in efficiency for those waveforms.
The EOB waveforms, on the other hand, extend beyond
the ISCO and thus have more cycles in the LIGO band than the TaylorT3 or the 
Padi\'eT1 waveforms. 
The power of the signal in band is sufficient for those waveforms to be
detected with an approximately equal probability for the
higher masses.


\section{Upper limit on the rate of BBH inspirals}
\label{interpretation}

As was mentioned previously, a reliable upper limit cannot be calculated
for the rate of BBH inspirals as was calculated for the rate of 
BNS inspirals. The reason for this is two-fold. Firstly,
the characteristics of the BBH population 
(such as spatial, mass and spin distributions) are not known, since no BBH
systems have ever been observed. 
In addition, the BCV templates are not guaranteed to have a good overlap
with the true BBH inspiral gravitational wave signals. 
However, work by various groups has given insights on the possible spatial
distribution and mass distribution of BBH systems
\cite{Belczynski:2002}. 
Assuming that the model-based inspiral waveforms proposed in the
literature have good overlap with a true inspiral gravitational wave
signal and because the
BCV templates have a good overlap with those model-based BBH inspiral 
waveforms, we
used some of those predictions to give an interpretation of the result of our
search.

\subsection{Upper limit calculation}

Following the notation used in \cite{LIGOS1iul}, let $\mathcal{R}$
indicate the rate of BBH coalescences per year per
Milky Way Equivalent Galaxy (MWEG) and $N_\mathrm{G}(\rho^\ast)$ indicate the
number of MWEGs which our search probes at $\rho \geq
\rho^\ast$.  The probability of observing an inspiral signal with
$\rho > \rho^\ast$ in an observation time $T$ is
\begin{equation}
  P(\rho>\rho^\ast;{\mathcal{R}}) = 1 - e^{-{\mathcal{R}}T 
  N_\mathrm{G}(\rho^\ast)}.
  \label{e:foreground-poisson}
\end{equation}
A trigger can arise from either an inspiral signal in the data or from
background.   If $P_b$ denotes the probability that all background triggers
have SNR less than $\rho^\ast$,  then the probability of observing
one or more triggers with $\rho > \rho^\ast$ is 
\begin{equation}
  P(\rho>\rho^\ast;{\mathcal{R}},b) = 1 - 
  P_b e^{-{\mathcal{R}}T N_\mathrm{G}(\rho^\ast)}.  
\label{e:joint-dist}
\end{equation}

Given the probability $P_b$, the total observation time $T$, the
SNR of the loudest event $\rho_{\mathrm{max}}$, and
the number of MWEGs $N_{\mathrm{G}}(\rho_{\mathrm{max}})$
to which the search is sensitive, we find that a frequentist upper limit,
at 90\% confidence level, is 
\begin{equation}
  {\cal R} < {\cal R}_{90\%} =
  \frac{2.303+\ln P_b} {T N_{\mathrm{G}}(\rho_{\mathrm{max}})}.
\label{e:ul}
\end{equation}
For ${\mathcal{R}}>{\mathcal{R}}_{90\%}$, there is more than $90\%$
probability that at least one event would be observed with SNR greater
than $\rho_{\mathrm{max}}$.   Details of this method of determining an
upper limit can be found in \cite{loudestGWDAW03}.  In
particular, one obtains a conservative upper limit by setting
$P_b=1$.  We adopt this approach below because of uncertainties in our
background estimate.

During the 350.4 hours of non-playground data used in this search,
the highest combined SNR that was observed was 16.056.
The number $N_{\mathrm{G}}$ can be calculated, as in the BNS search
\cite{LIGOS2iul}, using the Monte-Carlo simulations that were performed.
The difference from the BNS search is that the injected signals were
not drawn from an astrophysical  population, but from
a population that assumes a uniform $\log$(distance) distribution.
The way this difference is handled is explained below.

Our model for the BBH population carried the following assumptions:

(1) Black holes of mass between 3 and 20 $M_\odot$ 
result exclusively from the evolution of stars, so that our BBH sources are 
present only in galaxies.

(2) The field population of BBHs is distributed
amongst galaxies in proportion to the galaxies' blue light (as
was assumed for BNS systems \cite{LIGOS2iul}).

(3) The component mass distribution is uniform, with values ranging from 3 to
20 $M_\odot$ \cite{Belczynski:2002,Bulik:2004}.

(4) The component spins are negligibly small.

(5) The waveforms are an equal mixture of EOB, Pad\'eT1 and TaylorT3 waveforms.

(6) The sidereal times of the coalescences are distributed uniformly
throughout the S2 run.

Assumption (4) was made because the BCV templates used in this search
were not intended to capture the amplitude modulations of the 
gravitational waveforms expected to result from BBH systems with
spinning components. However, studies performed by BCV 
\cite{BuonannoChenVallisneri:2003b}
showed that the BCV templates do have high overlaps (90\% on average,
the average taken over one thousand initial spin orientations)
with waveforms of spinning BBH systems of comparable component masses.
Templates that are more suitable for detection of spinning BBH 
have been developed \cite{BuonannoChenVallisneri:2003b} (BCV2)
and are being used in a search for the inspiral of such binaries in the S3 
LIGO data.

Assumption (5) was probably the most ad-hoc assumption in our
upper limit calculation. Since this calculation was primarily intended to be
illustrative of how our results can be used to set an upper limit,
the mix of the waveforms was
chosen for simplicity.  It should be apparent how to modify the
calculation to fit a different population model.

\begin{figure}[ht]
\begin{center}
\includegraphics[width=\linewidth]{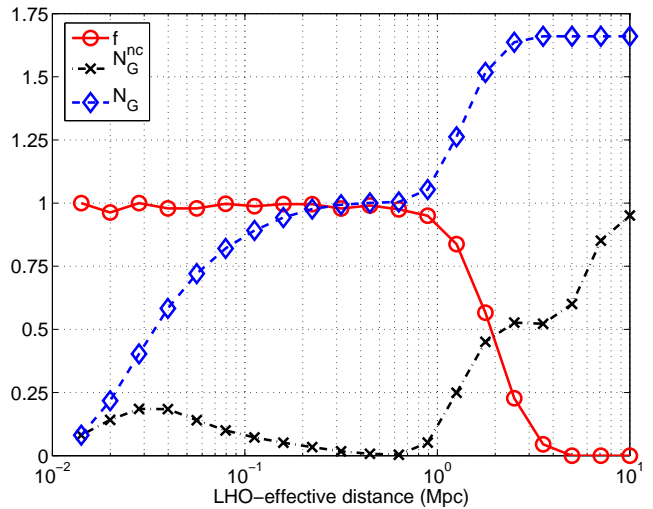}
\end{center}
\caption{\label{f:NG}
The number of Milky Way equivalent galaxies (crosses) and the efficiency of the
search (circles) as calculated by Eq. (\ref{finj}) versus the LHO-effective 
distance (in Mpc)
are shown. The cumulative number of MWEGs $N_\mathrm{G}$ (diamonds)
versus LHO effective distance is overlayed. 
The horizontal axis has a logarithmic scale, in accordance with the
uniform-$\log$ distance distribution of the injected BBH inspiral signals.
No units are given for the vertical axis 
because it corresponds to three different
quantities plotted against LHO-effective distance.}
\end{figure}

With assumptions (1) and (2) we determined the number of MWEGs 
in each logarithmic bin of 
LHO-effective distance ($N_\mathrm{G}^{\mathrm{nc}}$, crosses 
in Fig.~\ref{f:NG}).
We used our Monte-Carlo simulations to determine the efficiency for
detection of a source at LHO-effective distance $d$ with combined
SNR $\rho_\mathrm{C} > \rho_{\mathrm{max}}$, $\rho_{\mathrm{max}}$ 
being the combined SNR
of the loudest event observed in our search. Specifically, we define
\begin{eqnarray}
f(d; \rho_{\mathrm{max}}) &&=  \frac{1}{3} 
   \Big [ \frac{N_\mathrm{f}(d; \rho_{\mathrm{max}})}{N_{\mathrm{inj}}(d)} 
   \Big ]_{\mathrm{TaylorT3}} \\ \nonumber
   &&+  \frac{1}{3}\Big [ \frac{N_\mathrm{f}
   (d; \rho_{\mathrm{max}})}{N_{\mathrm{inj}}(d)} 
   \Big ]_{\mathrm{EOB}}
   + \frac{1}{3} \Big [ \frac{N_\mathrm{f}
   (d; \rho_{\mathrm{max}})}{N_{\mathrm{inj}(d)} }
   \Big ]_{\mathrm{PT1}}
\label{finj}
\end{eqnarray}
and that is also plotted in Fig.~\ref{f:NG} (circles). The efficiencies for
each waveform family individually are given in Table \ref{t:mwegs}.
Finally, $N_\mathrm{G}(\rho_{\mathrm{max}})$ was calculated as
\begin{equation}
N_\mathrm{G}(\rho_{\mathrm{max}}) = 
  \sum_{d = 0}^{\infty} f(d; \rho_{\mathrm{max}}) 
  \times N_\mathrm{G}^{\mathrm{nc}}(d).
\end{equation}
Evaluating that sum we obtained $N_\mathrm{G} = 1.6603$.
Using Eq.~(\ref{e:ul}) we obtained 
$\mathcal{R}_{90\%} = 35 \ \mathrm{year}^{-1} \ \mathrm{MWEG}^{-1}$.

\begin{table*}
\caption{\label{t:mwegs} Efficiencies of recovering simulated BBH inspiral
signals from three different waveform families for each LHO-effective distance
bin.}
\vspace*{0.1in}
\begin{ruledtabular}
\begin{tabular}{lccccccccc}
 & LHO-Effective Distance Range (Mpc)& 
 MWEGs & \multicolumn{3}{c}{Detected with $\rho>16.056$ versus Injected} & $N_{\mathrm{G}}$ \\ 
  &       &  & EOB & TaylorT3 & Pad\'eT1 & & \\
  & 0.0100 - 0.0141 & 0.0814 & 1 / 1  &   2 / 2 &   2 / 2  &     0.0814  &  \\
  & 0.0141 - 0.0200 & 0.1415 & 5 / 5  &   8 / 9 &   25 / 25  &   0.2177  &  \\
  & 0.0200 - 0.0282 & 0.1850 & 30 / 30  & 21 / 21 & 57 / 57  &  0.4027   &  \\
  & 0.0282 - 0.0398 & 0.1844 & 44 / 44  & 37 / 39 & 85 / 86  &  0.5832   &  \\
  & 0.0398 - 0.0562 & 0.1404 & 58 / 58  & 58 / 60 & 100 / 103  & 0.7207  &  \\
  & 0.0562 - 0.0794 & 0.0999 & 81 / 81  & 77 / 77 & 117 / 118  & 0.8203  &  \\
  & 0.0794 - 0.1122 & 0.0722 & 61 / 61  & 78 / 81 & 146 / 146  & 0.8916  &  \\
  & 0.1122 - 0.1585 & 0.0515 & 64 / 64  & 81 / 82 & 155 / 155  & 0.9429  &  \\
  & 0.1585 - 0.2239 & 0.0338 & 68 / 68  & 76 / 77 & 144 / 144  & 0.9766  &  \\
  & 0.2239 - 0.3162 & 0.0172 & 83 / 83  & 64 / 67 & 152 / 155  & 0.9934  &  \\
  & 0.3162 - 0.4467 & 0.0075 & 82 / 82  & 66 / 67 & 150 / 152  & 1.0009  &  \\
  & 0.4467 - 0.6310 & 0.0038 & 77 / 77  & 90 / 92 & 126 / 133  & 1.0046  &  \\
  & 0.6310 - 0.8913 & 0.0516 & 74 / 74  & 61 / 64 & 148 / 165  & 1.0536  &  \\
  & 0.8913 - 1.2589 & 0.2490 & 53 / 55  & 81 / 101 & 115 / 154  & 1.2622 &  \\
  & 1.2589 - 1.7783 & 0.4505 & 68 / 85  & 38 / 75 & 61 / 156  &  1.5171  &  \\
  & 1.7783 - 2.5119 & 0.5267 & 28 / 67  & 14 / 74 & 11 / 147  & 1.6368   &  \\
  & 2.5119 - 3.5481 & 0.5222 & 10 / 83  & 1 / 69 &  0 / 133  & 1.6603  &  \\
  & 3.5481 - 5.0119 & 0.6005 & 0 / 88  &  0 / 67 &  0 /156  &   1.6603  &  \\
  & 5.0119 - 7.0795 & 0.8510 & 0 / 83 &    0 / 65 &  0 /170  &   1.6603  &  \\
  & 7.0795 - 10.000 & 0.9507 & 0 / 71 &    0 / 65 &  0 /150  &   1.6603  &  \\
\end{tabular}
\end{ruledtabular}
\end{table*}

\subsection{Error analysis}

A detailed error analysis is necessary and was carefully done for the rate 
upper limit calculated for BNS inspirals in \cite{LIGOS2iul}. For the upper 
limit calculated here, such a detailed analysis was not possible due to the 
lack of a reliable BBH population. However, we estimated the errors coming from
calibration uncertainties and the errors due to the limited number of 
injections performed.

The effect of the calibration uncertainties was calculated as was
done in Sec.~IX A 2 of \cite{LIGOS2iul}. In principle, those uncertainties 
affect the combined SNR $\rho_\mathrm{C}$ as
\begin{equation}
\delta \rho_\mathrm{C} \le \max \Big 
\{ \Big [ \frac{\rho_{\mathrm{L}}^2}{\rho_\mathrm{C}^2}
   (\delta \rho_{\mathrm{L}})^2+ 
   \frac{\rho_{\mathrm{H}}^2}{\rho_\mathrm{C}^2}(\delta \rho_{\mathrm{H}})^2
   \Big ]^{1/2}, \
   2 (\delta \rho_{\mathrm{H}}), \ 2 (\delta \rho_{\mathrm{L}}) \Big \}
\end{equation}
where we modified Eq.~(23) of \cite{LIGOS2iul} based on our 
Eq.~(\ref{e:combsnr}) for the combined SNR. 
However, for the calculation of the effect of this error on the rate
upper limit, we were interested in how the calibration uncertainties would 
affect the combined SNR of the loudest event. 
Careful examination of the combined SNR of the loudest
event showed that for that event the minimum of the three possible values in 
Eq.~(\ref{e:combsnr}) was the value $(\rho_{\mathrm{L}}^2 + \rho_{\mathrm{H}}^2 
)^{1/2}$, so we calculated the error due to calibration uncertainties by
\begin{equation}
\delta \rho_\mathrm{C} \le 
 \Big [ \frac{\rho_{\mathrm{L}}^2}{\rho_\mathrm{C}^2}
   (\delta \rho_{\mathrm{L}})^2+
   \frac{\rho_{\mathrm{H}}^2}{\rho_\mathrm{C}^2}(\delta \rho_{\mathrm{H}})^2
   \Big ]^{1/2}.
\end{equation}
We simplified the calculation by being more conservative and using
\begin{equation}
\delta \rho_\mathrm{C} \le \Big [ (\delta \rho_{\mathrm{L}})^2 +
                    (\delta \rho_{\mathrm{H}})^2 \Big ]^{1/2} .
\end{equation}
We also used the fact that the maximum calibration errors at each site 
were $8.5\%$
for L1 and $4.5\%$ for LHO (as explained in \cite{LIGOS2iul}) to obtain
\begin{equation}
\delta \rho_\mathrm{C} \le \Big [ (0.085 \ \rho_{\mathrm{L}}^{\mathrm{max}})^2+
           (0.045 \rho_{\mathrm{H}}^{\mathrm{max}})^2 \Big ]^{1/2}.
\end{equation}
The resulting error in $N_\mathrm{G}$ was
\begin{equation}
\delta N_\mathrm{G} |_{\mathrm{cal}} = \pm 0.0859 \ \mathrm{MWEG}.
\end{equation}

The errors due to the limited number of injections in our Monte-Carlo 
simulations
had to be calculated for each logarithmic distance bin and the
resulting errors to be combined in quadrature. Specifically,
\begin{equation}
\delta N_\mathrm{G} |_{\mathrm{MC}} = \Big \{ \sum_{d = 0}^{\infty} 
\Big [ \delta f(d; \rho_{\mathrm{max}}) 
\times N_\mathrm{G}^{\mathrm{nc}}(d) \Big ]^2 \Big \}^{1/2}.
\end{equation}
Because $f$ was calculated using the 3 different waveform families, the
error $\delta f(d; \rho_{\mathrm{max}})$ is
\begin{equation}
  \delta f(d; \rho_{\mathrm{max}}) = \Big [ \sum_{\mathrm{appr}} 
  \frac{N_\mathrm{f}(d; \rho_{\mathrm{max}})  
  (N_{\mathrm{inj}}(d) -N_\mathrm{f}
(d; \rho_{\mathrm{max}}))}{ 3^2 \ [N_{\mathrm{inj}}(d)]^3}  \Big ]^{1/2}
\end{equation}
where the sum was calculated over the three waveform families: 
EOB, Pad\'eT1 and TaylorT3. This gave 
\begin{equation}
\delta N_\mathrm{G} |_{\mathrm{MC}} = \pm 0.0211 \ \mathrm{MWEG}.
\end{equation}

Finally we added the errors in $N_\mathrm{G}$ in quadrature and obtained 
\begin{equation}
\delta N_\mathrm{G} = \pm 0.0885 \ \mathrm{MWEG}.
\end{equation}
Both contributions to this error can be thought of a 1-$\sigma$ variations.
In order to calculate the 90\% level of the systematic errors we multiplied
$\delta N_\mathrm{G}$ by 1.6, so
\begin{equation}
\delta N_\mathrm{G} |_{90\%} = \pm 0.1415 \ \mathrm{MWEG}.
\end{equation}
To be conservative, we assumed a downward excursion 
$N_\mathrm{G} = (1.6603 - 0.1415) \ \mathrm{MWEG}
 = 1.5188 \ \mathrm{MWEG}$. When substituted
into the rate upper limit equation this gave 
\begin{equation}
{\cal R}_{90\%} = 38 \ \mathrm{year}^{-1} \ \mathrm{MWEG}^{-1}.
\end{equation}

\section{Conclusions and future prospects}
\label{conclusions}

We performed the first search for binary black hole inspiral signals in
data from the LIGO interferometers. This search, even though similar in
some ways to the binary neutron star inspiral search,
has some significant differences and presents unique challenges.
There were no events that could be indentified as gravitational waves.

The fact that the performance and sensitivity of the LIGO interferometers
is improving and the frequency sensitivity band is being extended to lower
frequencies makes us hopeful that the first detection of gravitational
waves from the inspiral phase of binary black hole coalescences may
happen in the near future.
In the absense of a detection, astrophysically interesting results
can be expected by LIGO very soon. The current most optimistic rates
for BBH coalescences are of the order of $10^{-4} \; \mathrm{year}^{-1}
\;\mathrm{MWEG}^{-1}$ \cite{OShaughnessy:2005}. It is estimated that at design
sensitivity the LIGO interferometers will be able to detect binary black hole
inspirals in at least 5600 MWEGs with
the most optimistic calculations giving up to 13600 MWEGs
\cite{Nutzman:2004}. A science run of two years at design sensitivity is
expected to give BBH coalescence rate upper limits of less than $10^{-4} \; 
\mathrm{year}^{-1} \; \mathrm{MWEG}^{-1}$.

\appendix
\section{Filtering details}
\label{appendixA}

The amplitude part of the BCV templates $\tilde{h}(f)$ can be decomposed into 
two pieces, which are linear
combinations of $f^{-7/6}$ and $f^{-1/2}$. Those expressions can be used to
construct an orthogonal basis ${\hat{h}_j}$ for the 4-dimensional linear
subspace of templates with $\phi_0 \in \left[0, 2\pi\right)$ and
$\alpha \in \left(-\infty,+\infty \right)$.
Specifically, we want the basis vectors to satisfy
\begin{equation}
\left < \hat{h}_k , \hat{h}_j \right > = \delta_{kj}.
\label{OrthonormBasis}
\end{equation}
To do this we construct two real functions $A_1(f)$ and $A_2(f)$, linear
combinations of $f^{-7/6}$ and $f^{-1/2}$, which are related to the four basis
vectors via:
\begin{eqnarray}
 \hat{h}_{1,2}(f) &=& A_{1,2}(f) \: e^{i \psi_s(f)} \\
 \hat{h}_{3,4}(f) &=& A_{1,2}(f) \: i \: e^{i \psi_s(f)}.
\end{eqnarray}
Then, Eq. (\ref{OrthonormBasis}) becomes:
\begin{equation}
4 \Re \int_0^{\infty} \frac{A_k(f) A_j(f)}{S_h(f)} df = \delta_{kj}.
\label{OrthonormA}
\end{equation}
Since the templates have to be normalized, $A_1(f)$ and $A_2(f)$ must satisfy
\[
\left[
 \begin{array}{c}
 A_1(f)\\
 A_2(f)
\end{array}
\right]
=\left[
 \begin{array}{cc}
 a_{1} & 0\\
 b_{1} & b_{2}
\end{array}\right]
\left[
 \begin{array}{c}
 f^{-7/6} \\
 f^{-1/2}
\end{array}\right].
\]
Imposing condition (\ref{OrthonormA}) gives the numerical values of the
normalization factors. Those are
\begin{eqnarray}
a_1 &=& I_{7/3}^{-1/2}, \\
b_1 &=& - \frac{I_{5/3}}{I_{7/3}} \big ( I_1 - \frac{I_{5/3}^2}{I_{7/3}}
  \big ) ^{-1/2}, \\
b_2 &=& \big ( I_1 - \frac{I_{5/3}^2}{I_{7/3}}
  \big ) ^{-1/2}, \\
\end{eqnarray}
where the integrals $I_k$ are
\begin{equation}
I_k = 4 \int_0^{f_{\mathrm{cut}}} \frac{df}{f^k S_h(f)}.
\end{equation}

The next step is to write the normalized template in terms of the 4 basis
vectors
\begin{equation}
\hat{h}(f) = c_1 \hat{h}_1(f) + c_2 \hat{h}_2(f)
        + c_3 \hat{h}_3(f) + c_4 \hat{h}_4(f)
\end{equation}
with
\begin{eqnarray}
c_1 &=& \cos\phi_0 \cos\omega, \\
c_2 &=& \cos\phi_0 \sin\omega, \\
c_3 &=& \sin\phi_0 \cos\omega, \\
c_4 &=& \sin\phi_0 \sin\omega,
\end{eqnarray}
where $\omega$ is related to $\alpha$ by
\begin{equation}
\tan \omega = -\frac{a_{1} \alpha}{b_{2}+b_{1}\alpha}.
\end{equation}

Once the filters are designed, the overlap is calculated and is equal to
\begin{eqnarray}
\label{overlap2}
\rho = \left <{s}, \hat{h}\right >  &=& K_1 \cos \omega \cos \phi_0
 + K_2 \sin \omega \cos \phi_0 \\ 
&&+ K_3 \cos \omega \sin \phi_0 + K_4 \sin \omega \sin \phi_0 \nonumber
\end{eqnarray}
where $K_j = \left < {s}, \hat{h}_j\right >, k=1,2,3,4$ are the
four integrals that are necessary, namely
\begin{equation}
K_1 = \Re
\int_0^{f_{\mathrm{cut}}} \frac{4 \tilde{s}(f) a_1 f^{-7/6} }
{S_h(f)} e^{-i \psi_s(f)} df , 
\end{equation}
\begin{equation}
K_2 = \Re \int_0^{f_{\mathrm{cut}}} \frac{4 \tilde{s}(f)
(b_1 f^{-7/6} + b_2 f^{-1/2}) }{S_h(f)}e^{-i \psi_s(f)} df  ,
\end{equation}
\begin{equation}
K_3 = -\Im \int_0^{f_{\mathrm{cut}}} \frac{4 \tilde{s}(f) a_1 f^{-7/6}}
{S_h(f)} e^{-i \psi_s(f)} df  ,
\end{equation}
\begin{equation}
K_4 = -\Im \int_0^{f_{\mathrm{cut}}} \frac{4 \tilde{s}(f)
(b_1 f^{-7/6} + b_2 f^{-1/2})}{ S_h(f)} e^{-i \psi_s(f)} df.
\end{equation}
Maximizing the SNR over $\phi_0$ and $\omega$ we get
\begin{eqnarray}
\rho_{\mathrm{maximized}} &= \frac{1}{2}
\sqrt{(K_1+K_4)^2 + (K_2-K_3)^2} \\
+& \frac{1}{2} \sqrt{(K_1-K_4)^2 + (K_2+K_3)^2}.
\nonumber
\label{e:rhomax}
\end{eqnarray}
The values of $\phi_0$ and $\alpha$ that give the maximized SNR are
\begin{equation}
\phi_0^{\mathrm{max}} = \frac{1}{2} \arctan{ \frac{K_2 + K_3}{K_1-K_4} }
- \frac{1}{2} \arctan{ \frac{K_2-K_3}{K_1+K_4} },
\end{equation}
\begin{equation}
\alpha^{\mathrm{max}} = -
\frac{b_2 \tan{\omega^{\mathrm{max}}}}{a_1 + b_1 \tan{\omega^{\mathrm{max}}}},
\end{equation}
where
\begin{equation}
\omega^{\mathrm{max}} = \frac{1}{2} \arctan{ \frac{K_2 - K_3}{K_1+K_4} }
+ \frac{1}{2} \arctan{ \frac{K_2+K_3}{K_1-K_4}}.
\end{equation}
An extensive discussion on the values of $\alpha$ is provided in 
Sec.~\ref{sss:singleifotuning}.

An equivalent expression for the SNR,
which is computationally less costly, can be produced if we define
\begin{eqnarray}
&F_1 = K_1 - i K_3, \\
&F_2 = K_2 - i K_4.  \nonumber
\end{eqnarray}
Eq.~\ref{e:rhomax} can then be written as
\begin{eqnarray}
\rho_{\mathrm{maximized}} &=
 \frac{1}{2} \sqrt{|F_1|^2 +|F_2|^2+2\Im (F_1 F_2^{\ast})} \\
&+\frac{1}{2} \sqrt{|F_1|^2 +|F_2|^2 -2\Im (F_1 F_2^{\ast})}.
\nonumber
\end{eqnarray}

\acknowledgments
The authors gratefully acknowledge the support of the United States National 
Science Foundation for the construction and operation of the LIGO Laboratory 
and the Particle Physics and Astronomy Research Council of the United Kingdom, 
the Max-Planck-Society and the State of Niedersachsen/Germany for support of 
the construction and operation of the GEO600 detector. The authors also 
gratefully acknowledge the support of the research by these agencies and by the 
Australian Research Council, the Natural Sciences and Engineering Research 
Council of Canada, the Council of Scientific and Industrial Research of India, 
the Department of Science and Technology of India, the Spanish Ministerio de 
Educacion y Ciencia, the John Simon Guggenheim Foundation, the Leverhulme Trust,
 the David and 
Lucile Packard Foundation, the Research Corporation, and the Alfred P. Sloan 
Foundation.


\bibliography{paper_archive}

\begin{thebibliography}{35}
\expandafter\ifx\csname natexlab\endcsname\relax\def\natexlab#1{#1}\fi
\expandafter\ifx\csname bibnamefont\endcsname\relax
  \def\bibnamefont#1{#1}\fi
\expandafter\ifx\csname bibfnamefont\endcsname\relax
  \def\bibfnamefont#1{#1}\fi
\expandafter\ifx\csname citenamefont\endcsname\relax
  \def\citenamefont#1{#1}\fi
\expandafter\ifx\csname url\endcsname\relax
  \def\url#1{\texttt{#1}}\fi
\expandafter\ifx\csname urlprefix\endcsname\relax\def\urlprefix{URL }\fi
\providecommand{\bibinfo}[2]{#2}
\providecommand{\eprint}[2][]{\url{#2}}

\bibitem[{\citenamefont{Abramovici et~al.}(1992)}]{Abramovici:1992ah}
\bibinfo{author}{\bibfnamefont{A.}~\bibnamefont{Abramovici}}
  \bibnamefont{et~al.}, \bibinfo{journal}{Science}
  \textbf{\bibinfo{volume}{256}}, \bibinfo{pages}{325} (\bibinfo{year}{1992}).

\bibitem[{\citenamefont{Abbott et~al.}(2005)}]{LIGOS2iul}
\bibinfo{author}{\bibfnamefont{B.}~\bibnamefont{Abbott}} \bibnamefont{et~al.}
  (\bibinfo{collaboration}{LIGO Scientific Collaboration}),
  \bibinfo{journal}{submitted to Phys. Rev. D}  (\bibinfo{year}{2005}),
  \eprint{web location:gr-qc/0505041}.

\bibitem[{\citenamefont{Abbott et~al.}(2004{\natexlab{a}})}]{LIGObursts:2004}
\bibinfo{author}{\bibfnamefont{B.}~\bibnamefont{Abbott}} \bibnamefont{et~al.},
  \bibinfo{journal}{Class. Quant. Grav.} \textbf{\bibinfo{volume}{21}},
  \bibinfo{pages}{S677} (\bibinfo{year}{2004}{\natexlab{a}}).

\bibitem[{\citenamefont{Damour et~al.}(1998)\citenamefont{Damour, Iyer, and
  Sathyaprakash}}]{Damour:1998zb}
\bibinfo{author}{\bibfnamefont{T.}~\bibnamefont{Damour}},
  \bibinfo{author}{\bibfnamefont{B.~R.} \bibnamefont{Iyer}}, \bibnamefont{and}
  \bibinfo{author}{\bibfnamefont{B.~S.} \bibnamefont{Sathyaprakash}},
  \bibinfo{journal}{Phys. Rev. D} \textbf{\bibinfo{volume}{57}},
  \bibinfo{pages}{885} (\bibinfo{year}{1998}).

\bibitem[{\citenamefont{Buonanno and Damour}(1999)}]{BuonannoDamour:1999}
\bibinfo{author}{\bibfnamefont{A.}~\bibnamefont{Buonanno}} \bibnamefont{and}
  \bibinfo{author}{\bibfnamefont{T.}~\bibnamefont{Damour}},
  \bibinfo{journal}{Phys. Rev. D} \textbf{\bibinfo{volume}{59}},
  \bibinfo{pages}{084006} (\bibinfo{year}{1999}).

\bibitem[{\citenamefont{Buonanno and Damour}(2000)}]{BuonannoDamour:2000}
\bibinfo{author}{\bibfnamefont{A.}~\bibnamefont{Buonanno}} \bibnamefont{and}
  \bibinfo{author}{\bibfnamefont{T.}~\bibnamefont{Damour}},
  \bibinfo{journal}{Phys. Rev. D} \textbf{\bibinfo{volume}{62}},
  \bibinfo{pages}{064015} (\bibinfo{year}{2000}).

\bibitem[{\citenamefont{Damour et~al.}(2000)\citenamefont{Damour, Jaranowski,
  and Sch{\"a}fer}}]{DamourJaranowskiSchaefer:2000}
\bibinfo{author}{\bibfnamefont{T.}~\bibnamefont{Damour}},
  \bibinfo{author}{\bibfnamefont{P.}~\bibnamefont{Jaranowski}},
  \bibnamefont{and}
  \bibinfo{author}{\bibfnamefont{G.}~\bibnamefont{Sch{\"a}fer}},
  \bibinfo{journal}{Phys. Rev. D} \textbf{\bibinfo{volume}{62}},
  \bibinfo{pages}{084011} (\bibinfo{year}{2000}).

\bibitem[{\citenamefont{Damour et~al.}(2001)\citenamefont{Damour, Iyer, and
  Sathyaprakash}}]{Damour:2000zb}
\bibinfo{author}{\bibfnamefont{T.}~\bibnamefont{Damour}},
  \bibinfo{author}{\bibfnamefont{B.~R.} \bibnamefont{Iyer}}, \bibnamefont{and}
  \bibinfo{author}{\bibfnamefont{B.~S.} \bibnamefont{Sathyaprakash}},
  \bibinfo{journal}{Phys. Rev. D} \textbf{\bibinfo{volume}{63}},
  \bibinfo{pages}{044023} (\bibinfo{year}{2001}).

\bibitem[{\citenamefont{Damour et~al.}(2003)\citenamefont{Damour, Iyer,
  Jaranowski, and Sathyaprakash}}]{DamouretAl:2003}
\bibinfo{author}{\bibfnamefont{T.}~\bibnamefont{Damour}},
  \bibinfo{author}{\bibfnamefont{B.}~\bibnamefont{Iyer}},
  \bibinfo{author}{\bibfnamefont{P.}~\bibnamefont{Jaranowski}},
  \bibnamefont{and} \bibinfo{author}{\bibfnamefont{B.~S.}
  \bibnamefont{Sathyaprakash}}, \bibinfo{journal}{Phys. Rev. D}
  \textbf{\bibinfo{volume}{67}}, \bibinfo{pages}{064028}
  (\bibinfo{year}{2003}).

\bibitem[{\citenamefont{Blanchet et~al.}(2004)\citenamefont{Blanchet, Damour,
  Esposito-Farese, and Iyer}}]{Blanchet:2004ek}
\bibinfo{author}{\bibfnamefont{L.}~\bibnamefont{Blanchet}},
  \bibinfo{author}{\bibfnamefont{T.}~\bibnamefont{Damour}},
  \bibinfo{author}{\bibfnamefont{G.}~\bibnamefont{Esposito-Farese}},
  \bibnamefont{and} \bibinfo{author}{\bibfnamefont{B.~R.} \bibnamefont{Iyer}},
  \bibinfo{journal}{Phys. Rev. Lett.} \textbf{\bibinfo{volume}{93}},
  \bibinfo{pages}{091101} (\bibinfo{year}{2004}), \eprint{gr-qc/0406012}.

\bibitem[{\citenamefont{Echeverria}(1989)}]{Echeverria:1989}
\bibinfo{author}{\bibfnamefont{F.}~\bibnamefont{Echeverria}},
  \bibinfo{journal}{Phys. Rev. D} \textbf{\bibinfo{volume}{40}},
  \bibinfo{pages}{3194} (\bibinfo{year}{1989}).

\bibitem[{\citenamefont{Leaver}(1985)}]{Leaver:1985}
\bibinfo{author}{\bibfnamefont{E.~W.} \bibnamefont{Leaver}},
  \bibinfo{journal}{Proc. R. Soc. London Ser. A}
  \textbf{\bibinfo{volume}{402}}, \bibinfo{pages}{285} (\bibinfo{year}{1985}).

\bibitem[{\citenamefont{Creighton}(1999)}]{Jolien:1999}
\bibinfo{author}{\bibfnamefont{J.~D.~E.} \bibnamefont{Creighton}},
  \bibinfo{journal}{Phys. Rev. D} \textbf{\bibinfo{volume}{60}},
  \bibinfo{pages}{022001} (\bibinfo{year}{1999}).

\bibitem[{\citenamefont{Blanchet et~al.}(2002)\citenamefont{Blanchet, Faye,
  Iyer, and Joguet}}]{Blanchet:2001ax}
\bibinfo{author}{\bibfnamefont{L.}~\bibnamefont{Blanchet}},
  \bibinfo{author}{\bibfnamefont{G.}~\bibnamefont{Faye}},
  \bibinfo{author}{\bibfnamefont{B.~R.} \bibnamefont{Iyer}}, \bibnamefont{and}
  \bibinfo{author}{\bibfnamefont{B.}~\bibnamefont{Joguet}},
  \bibinfo{journal}{Phys. Rev.} \textbf{\bibinfo{volume}{D65}},
  \bibinfo{pages}{061501} (\bibinfo{year}{2002}), \eprint{gr-qc/0105099}.

\bibitem[{\citenamefont{Blanchet}(2002)}]{Blanchet:2002av}
\bibinfo{author}{\bibfnamefont{L.}~\bibnamefont{Blanchet}},
  \bibinfo{journal}{Living Rev. Rel.} \textbf{\bibinfo{volume}{5}},
  \bibinfo{pages}{3} (\bibinfo{year}{2002}), \eprint{gr-qc/0202016}.

\bibitem[{\citenamefont{Blanchet et~al.}(1996)\citenamefont{Blanchet, Iyer,
  Will, and Wiseman}}]{Blanchet:1996pi}
\bibinfo{author}{\bibfnamefont{L.}~\bibnamefont{Blanchet}},
  \bibinfo{author}{\bibfnamefont{B.~R.} \bibnamefont{Iyer}},
  \bibinfo{author}{\bibfnamefont{C.~M.} \bibnamefont{Will}}, \bibnamefont{and}
  \bibinfo{author}{\bibfnamefont{A.~G.} \bibnamefont{Wiseman}},
  \bibinfo{journal}{Class. Quant. Grav.} \textbf{\bibinfo{volume}{13}},
  \bibinfo{pages}{575} (\bibinfo{year}{1996}).

\bibitem[{\citenamefont{Baker et~al.}(2002{\natexlab{a}})\citenamefont{Baker,
  Campanelli, Lousto, and Takahashi}}]{Baker:2002qf}
\bibinfo{author}{\bibfnamefont{J.~G.} \bibnamefont{Baker}},
  \bibinfo{author}{\bibfnamefont{M.}~\bibnamefont{Campanelli}},
  \bibinfo{author}{\bibfnamefont{C.~O.} \bibnamefont{Lousto}},
  \bibnamefont{and}
  \bibinfo{author}{\bibfnamefont{R.}~\bibnamefont{Takahashi}},
  \bibinfo{journal}{Phys. Rev. D} \textbf{\bibinfo{volume}{65}},
  \bibinfo{pages}{124012} (\bibinfo{year}{2002}{\natexlab{a}}),
  \eprint{astro-ph/0202469}.

\bibitem[{\citenamefont{Baker et~al.}(2002{\natexlab{b}})\citenamefont{Baker,
  Campanelli, and Lousto}}]{Baker:2001sf}
\bibinfo{author}{\bibfnamefont{J.~G.} \bibnamefont{Baker}},
  \bibinfo{author}{\bibfnamefont{M.}~\bibnamefont{Campanelli}},
  \bibnamefont{and} \bibinfo{author}{\bibfnamefont{C.~O.}
  \bibnamefont{Lousto}}, \bibinfo{journal}{Phys. Rev. D}
  \textbf{\bibinfo{volume}{65}}, \bibinfo{pages}{044001}
  (\bibinfo{year}{2002}{\natexlab{b}}), \eprint{gr-qc/0104063}.

\bibitem[{\citenamefont{Buonanno
  et~al.}(2003{\natexlab{a}})\citenamefont{Buonanno, Chen, and
  Vallisneri}}]{BuonannoChenVallisneri:2003a}
\bibinfo{author}{\bibfnamefont{A.}~\bibnamefont{Buonanno}},
  \bibinfo{author}{\bibfnamefont{Y.}~\bibnamefont{Chen}}, \bibnamefont{and}
  \bibinfo{author}{\bibfnamefont{M.}~\bibnamefont{Vallisneri}},
  \bibinfo{journal}{Phys. Rev. D} \textbf{\bibinfo{volume}{67}},
  \bibinfo{pages}{024016} (\bibinfo{year}{2003}{\natexlab{a}}).

\bibitem[{\citenamefont{Droz et~al.}(1999)\citenamefont{Droz, Knapp, Poisson,
  and Owen}}]{Droz:1999qx}
\bibinfo{author}{\bibfnamefont{S.}~\bibnamefont{Droz}},
  \bibinfo{author}{\bibfnamefont{D.~J.} \bibnamefont{Knapp}},
  \bibinfo{author}{\bibfnamefont{E.}~\bibnamefont{Poisson}}, \bibnamefont{and}
  \bibinfo{author}{\bibfnamefont{B.~J.} \bibnamefont{Owen}},
  \bibinfo{journal}{Phys. Rev. D} \textbf{\bibinfo{volume}{59}},
  \bibinfo{pages}{124016} (\bibinfo{year}{1999}).

\bibitem[{\citenamefont{Thorne}(1987)}]{thorne.k:1987}
\bibinfo{author}{\bibfnamefont{K.~S.} \bibnamefont{Thorne}}, in
  \emph{\bibinfo{booktitle}{Three hundred years of gravitation}}, edited by
  \bibinfo{editor}{\bibfnamefont{S.~W.} \bibnamefont{Hawking}}
  \bibnamefont{and} \bibinfo{editor}{\bibfnamefont{W.}~\bibnamefont{Israel}}
  (\bibinfo{publisher}{Cambridge University Press},
  \bibinfo{address}{Cambridge}, \bibinfo{year}{1987}),
  chap.~\bibinfo{chapter}{9}, pp. \bibinfo{pages}{330--458}.

\bibitem[{\citenamefont{Sathyaprakash and
  Dhurandhar}(1991)}]{SathyaDhurandhar:1991}
\bibinfo{author}{\bibfnamefont{B.~S.} \bibnamefont{Sathyaprakash}}
  \bibnamefont{and} \bibinfo{author}{\bibfnamefont{S.~V.}
  \bibnamefont{Dhurandhar}}, \bibinfo{journal}{Phys. Rev D}
  \textbf{\bibinfo{volume}{44}}, \bibinfo{pages}{3819} (\bibinfo{year}{1991}).

\bibitem[{\citenamefont{Fairhurst et~al.}(2004)}]{S2Hardware:2004}
\bibinfo{author}{\bibfnamefont{S.}~\bibnamefont{Fairhurst}}
  \bibnamefont{et~al.}, \emph{\bibinfo{title}{Analysis of inspiral hardware
  injections during s2}} (\bibinfo{year}{2004}), \bibinfo{note}{in
  preparation}.

\bibitem[{\citenamefont{Christensen et~al.}(2004)\citenamefont{Christensen,
  Shawhan, and Gonz{\'a}lez}}]{vetoGWDAW03}
\bibinfo{author}{\bibfnamefont{N.}~\bibnamefont{Christensen}},
  \bibinfo{author}{\bibfnamefont{P.}~\bibnamefont{Shawhan}}, \bibnamefont{and}
  \bibinfo{author}{\bibfnamefont{G.}~\bibnamefont{Gonz{\'a}lez}},
  \bibinfo{journal}{Class. Quant. Grav.} \textbf{\bibinfo{volume}{21}},
  \bibinfo{pages}{S1747} (\bibinfo{year}{2004}).

\bibitem[{\citenamefont{Allen et~al.}(2005)\citenamefont{Allen, Anderson,
  Brady, Brown, and Creighton}}]{findchirppaper}
\bibinfo{author}{\bibfnamefont{B.~A.} \bibnamefont{Allen}},
  \bibinfo{author}{\bibfnamefont{W.~G.} \bibnamefont{Anderson}},
  \bibinfo{author}{\bibfnamefont{P.~R.} \bibnamefont{Brady}},
  \bibinfo{author}{\bibfnamefont{D.~A.} \bibnamefont{Brown}}, \bibnamefont{and}
  \bibinfo{author}{\bibfnamefont{J.~D.~E.} \bibnamefont{Creighton}}
  (\bibinfo{year}{2005}), \eprint{gr-qc/0509116}.

\bibitem[{LAL()}]{LALS2BBH}
\emph{\bibinfo{title}{{\normalfont LSC Algorithm Library software packages
  {\scshape lal} and {\scshape lalapps}}}}, \bibinfo{note}{the CVS tag versions
  \texttt{inspiral-bcv-2004-10-28-0} and \texttt{inspiral-bcv-2005-02-03-0} of
  \textsc{lal} and \textsc{lalapps} were used in this analysis.},
  \urlprefix\url{http://www.lsc-group.phys.uwm.edu/lal}.

\bibitem[{\citenamefont{Allen}(2005)}]{Allen:2004}
\bibinfo{author}{\bibfnamefont{B.}~\bibnamefont{Allen}},
  \bibinfo{journal}{Phys. Rev. D} \textbf{\bibinfo{volume}{71}},
  \bibinfo{pages}{062001} (\bibinfo{year}{2005}).

\bibitem[{\citenamefont{Tannenbaum et~al.}(2001)\citenamefont{Tannenbaum,
  Wright, Miller, and Livny}}]{beowulfbook-condor}
\bibinfo{author}{\bibfnamefont{T.}~\bibnamefont{Tannenbaum}},
  \bibinfo{author}{\bibfnamefont{D.}~\bibnamefont{Wright}},
  \bibinfo{author}{\bibfnamefont{K.}~\bibnamefont{Miller}}, \bibnamefont{and}
  \bibinfo{author}{\bibfnamefont{M.}~\bibnamefont{Livny}}, in
  \emph{\bibinfo{booktitle}{Beowulf Cluster Computing with {L}inux}}, edited by
  \bibinfo{editor}{\bibfnamefont{T.}~\bibnamefont{Sterling}}
  (\bibinfo{publisher}{MIT Press}, \bibinfo{year}{2001}).

\bibitem[{\citenamefont{Belczynski et~al.}(2002)\citenamefont{Belczynski,
  Kalogera, and Bulik}}]{Belczynski:2002}
\bibinfo{author}{\bibfnamefont{K.}~\bibnamefont{Belczynski}},
  \bibinfo{author}{\bibfnamefont{V.}~\bibnamefont{Kalogera}}, \bibnamefont{and}
  \bibinfo{author}{\bibfnamefont{T.}~\bibnamefont{Bulik}},
  \bibinfo{journal}{Astrophys. J.} \textbf{\bibinfo{volume}{572}},
  \bibinfo{pages}{407} (\bibinfo{year}{2002}).

\bibitem[{\citenamefont{Abbott et~al.}(2004{\natexlab{b}})}]{LIGOS1iul}
\bibinfo{author}{\bibfnamefont{B.}~\bibnamefont{Abbott}} \bibnamefont{et~al.}
  (\bibinfo{collaboration}{LIGO Scientific Collaboration}),
  \bibinfo{journal}{Phys. Rev. D} \textbf{\bibinfo{volume}{69}},
  \bibinfo{pages}{122001} (\bibinfo{year}{2004}{\natexlab{b}}).

\bibitem[{\citenamefont{Brady et~al.}(2004)\citenamefont{Brady, Creighton, and
  Wiseman}}]{loudestGWDAW03}
\bibinfo{author}{\bibfnamefont{P.~R.} \bibnamefont{Brady}},
  \bibinfo{author}{\bibfnamefont{J.~D.~E.} \bibnamefont{Creighton}},
  \bibnamefont{and} \bibinfo{author}{\bibfnamefont{A.~G.}
  \bibnamefont{Wiseman}}, \bibinfo{journal}{Class. Quant. Grav.}
  \textbf{\bibinfo{volume}{21}}, \bibinfo{pages}{S1775} (\bibinfo{year}{2004}).

\bibitem[{\citenamefont{Bulik et~al.}(2004)\citenamefont{Bulik,
  Gondek-Rosinska, and Belczynski}}]{Bulik:2004}
\bibinfo{author}{\bibfnamefont{T.}~\bibnamefont{Bulik}},
  \bibinfo{author}{\bibfnamefont{D.}~\bibnamefont{Gondek-Rosinska}},
  \bibnamefont{and}
  \bibinfo{author}{\bibfnamefont{K.}~\bibnamefont{Belczynski}},
  \bibinfo{journal}{Mon. Not. Roy. Astron. Soc.}
  \textbf{\bibinfo{volume}{352}}, \bibinfo{pages}{1372} (\bibinfo{year}{2004}).

\bibitem[{\citenamefont{Buonanno
  et~al.}(2003{\natexlab{b}})\citenamefont{Buonanno, Chen, and
  Vallisneri}}]{BuonannoChenVallisneri:2003b}
\bibinfo{author}{\bibfnamefont{A.}~\bibnamefont{Buonanno}},
  \bibinfo{author}{\bibfnamefont{Y.}~\bibnamefont{Chen}}, \bibnamefont{and}
  \bibinfo{author}{\bibfnamefont{M.}~\bibnamefont{Vallisneri}},
  \bibinfo{journal}{Phys. Rev. D} \textbf{\bibinfo{volume}{67}},
  \bibinfo{pages}{104025} (\bibinfo{year}{2003}{\natexlab{b}}).

\bibitem[{\citenamefont{O'Shaughnessy et~al.}(2005)\citenamefont{O'Shaughnessy,
  Kim, Fragkos, Kalogera, and Belczynski}}]{OShaughnessy:2005}
\bibinfo{author}{\bibfnamefont{R.}~\bibnamefont{O'Shaughnessy}},
  \bibinfo{author}{\bibfnamefont{C.}~\bibnamefont{Kim}},
  \bibinfo{author}{\bibfnamefont{T.}~\bibnamefont{Fragkos}},
  \bibinfo{author}{\bibfnamefont{V.}~\bibnamefont{Kalogera}}, \bibnamefont{and}
  \bibinfo{author}{\bibfnamefont{K.}~\bibnamefont{Belczynski}},
  \bibinfo{journal}{ApJ, in press}  (\bibinfo{year}{2005}),
  \eprint{astro-ph/0504479}.

\bibitem[{\citenamefont{Nutzman et~al.}(2004)\citenamefont{Nutzman, Kalogera,
  Finn, Hendrickson, and Belczynski}}]{Nutzman:2004}
\bibinfo{author}{\bibfnamefont{P.}~\bibnamefont{Nutzman}},
  \bibinfo{author}{\bibfnamefont{V.}~\bibnamefont{Kalogera}},
  \bibinfo{author}{\bibfnamefont{L.~S.} \bibnamefont{Finn}},
  \bibinfo{author}{\bibfnamefont{C.}~\bibnamefont{Hendrickson}},
  \bibnamefont{and}
  \bibinfo{author}{\bibfnamefont{K.}~\bibnamefont{Belczynski}},
  \bibinfo{journal}{Astrophys. J.} \textbf{\bibinfo{volume}{612}},
  \bibinfo{pages}{364} (\bibinfo{year}{2004}).

\end{thebibliography}

\end{document}